\documentclass[longauth]{aa} 
\usepackage[nolist]{acronym}

\newcommand{\rhill}{$\mathrm{r_{Hill}}$} 

\newcommand{\bpb}{$\beta$\,Pic\,b}
\newcommand{\bp}{$\beta$\,Pic}

\usepackage{graphicx}
\usepackage{threeparttablex}
\usepackage{txfonts}
\usepackage{hyperref}
\hypersetup{colorlinks=true,allcolors=[rgb]{0,0,0.8}}

\graphicspath{{figs/}}

\makeatletter
\renewcommand*\aa@pageof{, page \thepage{} of \pageref*{LastPage}}
\makeatother

\newacro{cpd}[CPD]{circumplanetary disk}
\newacro{psf}[PSF]{point spread function}
\newacro{hst}[{\it HST}]{Hubble Space Telescope}

\begin{document}

\title{The $\beta$ Pictoris b Hill Sphere Transit Campaign}
\subtitle{I. Photometric limits to dust and rings}

\titlerunning{Dust and rings around Beta Pictoris b}
\authorrunning{Kenworthy et al.}

\author{M. A. Kenworthy\inst{1}
\and
S. N.~Mellon\inst{2}
\and
J. I. Bailey,~III\inst{3}
\and
R. Stuik\inst{1,4}
\and
P. Dorval\inst{1,4}
\and
G. J. J. Talens\inst{5}
\and
S.~R.~Crawford\inst{6,7}
\and
E.E.~Mamajek\inst{2,8}
\and
I.~Laginja\inst{9,10}
\and
M.~Ireland\inst{11}
\and
B.~Lomberg\inst{6,12,13}
\and
R.~B.~Kuhn\inst{6,14}
\and
I.~Snellen\inst{1}
\and
K.~Zwintz\inst{15}
\and
R.~Kuschnig\inst{16}
\and
G. M. Kennedy\inst{17,18}
\and
L. Abe\inst{19}
\and
A. Agabi\inst{19}
\and
D. Mekarnia\inst{19}
\and
T. Guillot\inst{19}
\and
F. Schmider\inst{19}
\and
P. Stee\inst{19}
\and
Y.~de~Pra\inst{20,21}
\and
M.~Buttu\inst{20}
\and
N.~Crouzet\inst{22}
\and
P.~Kalas\inst{23,24,25}
\and
J.~J.~Wang\inst{26}
\and
K.~Stevenson\inst{27,28}
\and
E.~de~Mooij\inst{29,30}
\and
A.-M.~Lagrange\inst{31,32,33}
\and
S.~Lacour\inst{32}
\and
A.~Lecavelier~des~Etangs\inst{34}
\and
M. Nowak\inst{32,35}
\and
P.~A.~Str\o{}m\inst{17}
\and
Z.~Hui\inst{36}
\and
L.~Wang\inst{37}
}

\institute{Leiden Observatory, Leiden University, Postbus 9513, 2300 RA Leiden, The Netherlands
 \and
Department of Physics \& Astronomy, University of Rochester, Rochester, NY 14627, USA
 \and
Department of Physics, University of California at Santa Barbara, Santa Barbara, CA 93106, USA
 \and
NOVA Optical IR Instrumentation Group at ASTRON, PO Box 2, 7990AA Dwingeloo, The Netherlands
 \and
Institut de Recherche sur les Exoplan\`{e}tes, D\'{e}partement de Physique, Universit\'{e} de Montr\'{e}al, Montr\'{e}al, QC H3C 3J7, Canada
 \and
South African Astronomical Observatory, Observatory Rd, Observatory Cape Town, 7700 Cape Town, South Africa
 \and
NASA Headquarters, 300 E Street SW, Washington, DC 20546, USA
 \and
Jet Propulsion Laboratory, California Institute of Technology, 4800 Oak Grove Drive, M/S321-100, Pasadena, CA 91109, USA
 \and
DOTA, ONERA, Universit\'e Paris Saclay, F-92322 Ch\^{a}tillon, France
 \and
Aix Marseille Universit\'{e}, CNRS, LAM (Laboratoire d'Astrophysique de Marseille) UMR 7326, 13388 Marseille, France
 \and
Research School of Astronomy and Astrophysics, Australian National University, Canberra, ACT 2611, Australia
 \and
Department of Astronomy, University of Cape Town, Rondebosch, 7700 Cape Town, South Africa
 \and
Astrofica Technologies Pty Ltd, 2 Francis Road, Zonnebloem, Woodstock, Cape town, 7925, South Africa
 \and
Southern African Large Telescope, Observatory Rd, Observatory Cape Town, 7700 Cape Town, South Africa
 \and
Institut f\"ur Astro- und Teilchenphysik, Universit\"at Innsbruck, Technikerstra{\ss}e 25, A-6020 Innsbruck
 \and
Institut f\"ur Kommunikationsnetze und Satellitenkommunikation, Technical University Graz, Inffeldgasse 12, A-8010 Graz, Austria
 \and
Department of Physics, University of Warwick, Coventry CV4 7AL, UK
 \and
Centre for Exoplanets and Habitability, University of Warwick, Gibbet Hill Road, Coventry CV4 7AL, UK
 \and
Universit\'{e} C\^{o}te d'Azur, Observatoire de la C\^{o}te d'Azur, CNRS, Laboratoire Lagrange, France
 \and
Concordia Station, IPEV/PNRA, Dome C, Antarctica
 \and
Department of Mathematics, Computer Science and Physics, University of Udine, Italy
 \and
European Space Agency (ESA), European Space Research and Technology Centre (ESTEC), Keplerlaan 1, 2201 AZ Noordwijk, The Netherlands
 \and
Astronomy Department, University of California, Berkeley, CA 94720, USA
 \and
SETI Institute, Carl Sagan Center, 189 Bernardo Ave.,  Mountain View CA 94043, USA
 \and
Institute of Astrophysics, FORTH, GR-71110 Heraklion, Greece
 \and
Department of Astronomy, California Institute of Technology, Pasadena, CA 91125, USA
 \and
Space Telescope Science Institute, Baltimore, MD 21218, USA
 \and
JHU Applied Physics Laboratory, 11100 Johns Hopkins Rd, Laurel, MD 20723, USA
 \and
Astrophysics Research Centre, Queen’s University Belfast, Belfast BT7 1NN, UK
 \and
School of Physical Sciences and Centre for Astrophysics \& Relativity, Dublin City University, Glasnevin, Dublin 9, Ireland
 \and
IPAG, Univ. Grenoble Alpes, CNRS, IPAG, F-38000 Grenoble, France
 \and
LESIA, Observatoire de Paris, Universit\'{e} PSL, CNRS, Sorbonne Universit\'{e}, Universit\'{e} de Paris, 5 place Jules Janssen, 92195 Meudon, France
 \and
IMCCE - Observatoire de Paris, 77 Avenue Denfert-Rochereau, F-75014 PARIS
 \and
Institut d’Astrophysique de Paris, UMR7095 CNRS, Universit\'{e} Pierre \& Marie Curie, 98 bis boulevard Arago, 75014 Paris, France
 \and
Institute of Astronomy, Madingley Road, Cambridge CB3 0HA, UK
 \and
Shanghai Observatory, Chinese Academy of Sciences, China
 \and
Purple Mountain Observatory, Chinese Academy of Science, Nanjing 210008, China
 }

\date{Received December 12, 2020; accepted February 8, 2021}

  \abstract
  {}
  {
    Photometric monitoring of \bp{} in 1981 showed anomalous fluctuations of up to 4\% over several days, consistent with foreground material transiting the stellar disk.
    The subsequent discovery of the gas giant planet \bpb{} and the predicted transit of its Hill sphere to within 0.1\,au of the planet provided an opportunity to search for the transit of a \ac{cpd} in this 21\,$\pm$\,4 Myr-old planetary system. 
We aim to detect or put an upper limit of the density and nature of the material in the circumplanetary environment of the planet through continuous photometric monitoring of the Hill sphere transit in 2017 and 2018.
  }
   {Continuous broadband photometric monitoring of \bp{} requires ground-based observatories at multiple longitudes to provide redundancy and to provide triggers for rapid spectroscopic followup. 
   These observatories include the dedicated \bp{} monitoring observatory bRing at Sutherland and Siding Springs, the ASTEP400 telescope at Concordia, and observations 
   from the space observatories BRITE and \ac{hst}.
   We search the combined light curves for evidence of short period transient events caused by rings and for longer term photometric variability due to diffuse circumplanetary material.}
   {We find no photometric event that matches with the event seen in November 1981, and there is no systematic photometric dimming of the star as a function of the Hill sphere radius.}
   {We conclude that the 1981 event was not caused by the transit of a \ac{cpd} around \bpb{}.
   The upper limit on the long term variability of \bp{} places an upper limit of $1.8\times 10^{22}$ g of dust within the Hill sphere (comparable to the $\sim$100\,km-radius asteroid 16 Psyche). 
   %
   %
   %
   Circumplanetary material is either condensed into a disk that does not transit \bp{}, is condensed into a disk with moons that has an obliquity that does not intersect with the path of \bp{} behind the Hill sphere, or is below our detection threshold.
   This is the first time that a dedicated international campaign has mapped the Hill sphere transit of a gas giant extrasolar planet at 10\,au.}
   

   \maketitle
%

\section{Introduction}

The formation of planetary systems is composed of several stages: the initial gravitational collapse of the prestellar cloud to form the protostar and a surrounding protostellar disk composed of gas and dust, the formation of protoplanetary cores within this circumstellar disk, and for the gas giant planets, the subsequent accretion of gas and dust onto the planet through a \acf{cpd} \citep{Lubow99, Lambrechts12, Mordasini18}.
When the protoplanetary disk disperses some $\sim$1-10 Myr after the birth of the star, the \ac{cpd} material subsequently accretes onto the young giant planets, spawns satellites, and then dissipates - likely through photoevaporation \citep[e.g.][]{Mamajek09, Canup02, Oberg20}.
We have strong evidence of the existence of circumplanetary disks in other planetary systems, notably hydrogen shocks seen from infalling gas onto the two planets in the PDS~70 \citep{Keppler18,Haffert19} system, and directly in sub-mm thermal emission with ALMA \citep{Isella19}.
The \ac{cpd} transitions from being optically thick with both gas and dust, through a phase where forming moons will create ring-like structures throughout the Hill sphere of the exoplanet before dispersing completely.
One such giant, transient exoring structure may have already been seen towards the young star J1407 \citep[V1400 Cen;][]{Mamajek12, Kenworthy15} and similar eclipsing events have been seen towards PDS~110 \citep{Osborn17,Osborn19} and the nearby star J0600 \citep{Way19,Way19b}.
The photometric fluctuations from the transit of the \ac{cpd} can be inverted into a radial map of the \ac{cpd}'s substructure and  indicate the location of moons in formation within them \citep{Kenworthy15}.

Additional \ac{cpd} transits can be discovered in wide field photometric surveys of star forming regions that contain planet forming systems, or by looking at known exoplanet systems with orbits of planets that are close to edge on from our line of sight.
The nearby, bright star \bp{} \citep[$d$\,=\,19.44\,pc,  $V$\,=\,3.85;][]{vanLeeuwen07b} has been intensively studied since the discovery and imaging of a nearly edge-on circumstellar debris disk \citep{Smith84,Kalas95} that extends out to 1800\,au.
A warp seen in the inner portion of the circumstellar disk \citep{Heap00}, combined with the detection of infalling comets \citep[see references in ][]{Kiefer14} implied the existence of at least one gas giant planet \citep{Mouillet97,Augereau01} which was discovered and confirmed by direct imaging \citep{Lagrange09,Lagrange10}.
Photometric \citep{Lous18} and spectroscopic transit searches \citep{vanSluijs19} did not reveal any transiting planets in the system, but more recently a second planet was detected through radial velocity monitoring of the star \citep{Lagrange19} and confirmed with observations with GRAVITY \citep{Nowak20,Lagrange20}.
The larger of the two planets, \bpb{}, is a gas giant planet with a mass of $\sim 11 M_{Jup}$ \citep{Lagrange20} and a highly inclined orbit that is close to edge on \citep{Millar-Blanchaer15,Wang16,Nielsen20,Lagrange20}.
The parameters of the star and \bpb{} are listed in Table~\ref{bpicparams}.
The star is a $\delta$ Scuti pulsator and shows millimagnitude variations on the timescale of 5 to 30 minutes \citep{koen2003a,koen2003b,Merkania17,Zwintz19}.
Stellar modelling and asteroseimology in \citet{Zwintz19} shows that the star rotates with $\sim$27\% Keplerian breakup velocity and has an inclination angle of 89.1 degrees (which matches with the inclination of the disk and planet b). 
A measurement of the planet's radial velocity by \citet{Snellen14} showed that the planet would move through inferior conjunction during the year 2017, and the orbital analysis by \citet{Wang16} showed that the planet would not transit the disk of the star, but that the star would pass within 20\% of the radius of the Hill sphere of \bpb{}.
More recent observations and analysis of the orbit of \bpb{} \citep{Lagrange19,Nielsen20} indicate that the impact parameter is closer to 10\% of the Hill sphere radius.
For a 11$M_{Jup}$ planet orbiting an 1.8 Solar mass star at 9.8\,au and $e\sim 0.09$ the radius of the Hill sphere is 1.1\,au.

%
%

This near transit provided a unique opportunity to monitor the circumplanetary environment of a young exoplanet, around one of the brightest known exoplanet host stars in the sky.
A workshop held in October 2016 brought several groups together to plan for the \bpb{} Hill sphere transit\footnote{The Lorentz Center workshop ``Rocks, Rubble and Rings'' held 25-30 September 2016 in Leiden, the Netherlands}.
Several photometric and spectroscopic observing campaigns were presented and coordinated, three of which were the bRing observatories in South Africa and Australia, the ASTEP 400 telescope in Antarctica, and one of the BRITE Constellation satellites.
The bRing observatories were specifically built to monitor the Hill sphere transit, providing longitudinal coverage of the star from two locations in the Southern hemisphere, and combined with data from the MASCARA South instrument commissioned in La Silla.
The ASTEP 400 telescope was developed for photometric transit searches during the Antarctic winters, and the BRITE Constellation satellites are used for precision photometric monitoring of pulsating stars and asteroseismology.
An observing campaign with the Hubble Space Telescope provided photometric calibration of the ground based data, and a space based cubesat called PicSat \citep{Nowak18} was built and launched to obtain dedicated monitoring of \bp. 
Unfortunately an issue with the communications of PicSat meant that it failed several weeks after it was launched, and the details are described in \citet{Nowak18}.

In this paper we present an analysis of the high cadence photometric monitoring campaigns from bRing, BRITE and ASTEP, and the observations from the \ac{hst}.
In Section~\ref{sec:obs} we describe the high cadence observations carried out with the three observatories, and then search these photometric time series for a transiting \ac{cpd} and for a repeat of the 1981 transit event seen towards the star.
Our discussion and conclusions in Section~\ref{sec:concl} covers the implications from our analysis, future observations and other \ac{cpd} transit searches.


\begin{table}
\centering
\begin{threeparttable}
\caption[Adopted Observational Values for the $\beta$ Pictoris System]{Adopted Observational Values for the $\beta$ Pictoris System}
\vspace{0.05in}
\label{bpicparams}
\begin{tabular}{|c |c |c| c|} 
\hline
Parameter\,&Value\,&Units\,&Reference\\
\hline
$M_*$       & 1.797 $\pm$ 0.035     & $M_\odot$ & 1\\
$R_*$       & 1.497 $\pm$ 0.025     & $R_\odot$ & 1\\
$T_*$       & 8090 $\pm$ 59         & K         & 1\\
$L_*$       & 8.47 $\pm$ 0.23     & $L_{\odot}$ & 1\\
$M_b$       & 11.1 $\pm$ 0.8      & $M_J$     & 2\\
$R_b$       & 1.46 $\pm$ 0.01     & $R_J$     & 3\\
$T_b$       & 1724 $\pm$ 15       & K         & 3\\
  a           & 9.76 $\pm$ 0.04 & AU       & 2 \\
  e          &   0.09 $\pm$ 0.01   &       & 2 \\
Age & 21 $\pm$ 4 & Myr & 4\\
\hline
\end{tabular}
\begin{tablenotes}
\small
\item (1) \citet{Zwintz19},
(2) \citet{Lagrange20},
(3) \citet{Chilcote17},
(4) adopted estimated which is consistent with the
combination of recent estimates based on 
kinematics \citep{Crundall19,Miret-Roig20}, Li depletion boundary \citep{Binks16,Shkolnik17}, and isochrones
\citep{Mamajek14,Bell15}. 
\end{tablenotes}
\end{threeparttable}
\end{table}

\section{Geometry of the Hill Sphere Transit}\label{sec:hs}

\begin{figure*}[htb]
\centering
\includegraphics[width=0.8\textwidth]{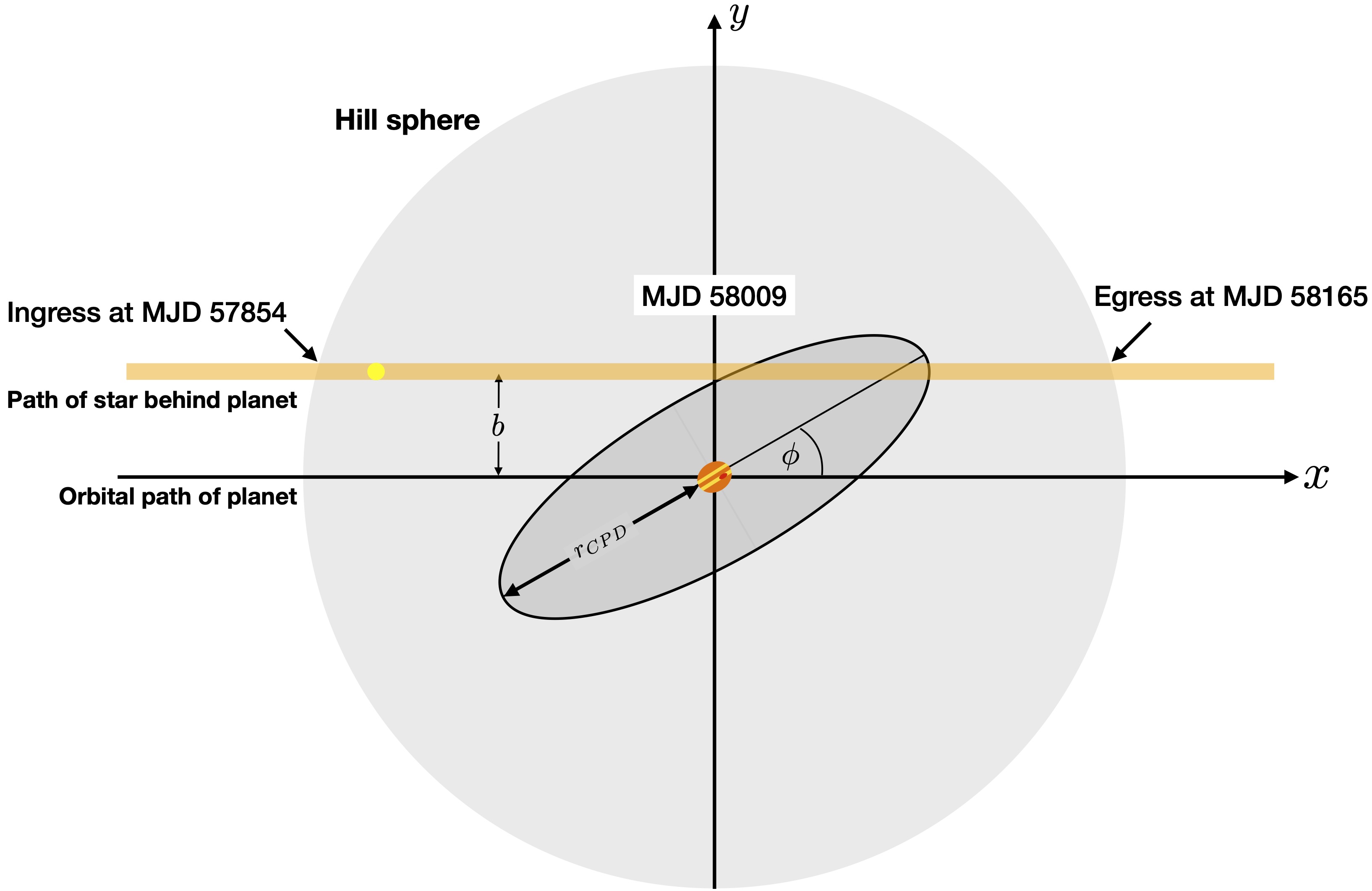}
\caption{Sketch of the circumplanetary disk model, showing how the coordinate system and orientation of the CPD is defined. The star moves on the defined path behind the Hill sphere and the CPD.}
\label{cpdmodel}
\end{figure*}

We adopt the values for the \bpb{} orbital parameters from ``NIRDIFS-GRAV-RV'' model of \citet{Lagrange20} and use them throughout the paper unless otherwise noted.
The transit of the \bpb{} Hill sphere takes approximately 311 days, with the Hill sphere ingress at 2017 Apr 11, midpoint of the transit at 2017 Sep 13 with a projected separation of star and planet of 0.11\,au, 9\%\, of the Hill sphere, and egress at 2018 Feb 16 - this is illustrated in Figure~\ref{cpdmodel} along with the dates in Modified Julian Dates.
These dates are indicated on plots of the time series in this paper with light and dark grey panels.
Even after recovering the position of the planet in 2018 \citep{Lagrange19}, there is an uncertainty of about 18 days for ingress and egress, and an error of 2.3 days on the day of closest approach.
The recent discovery and confirmation of the planet Beta Pictoris c \citep{Lagrange20,Nowak20} means that these dates vary slightly depending on the combination of astrometric measurements taken together, and whether the planets are constrained to be coplanar or not.
Any material at the orbital distance of the planet takes approximately 48 hours to cross the disk of the star.
To resolve any transits temporally therefore requires photometric monitoring on a timescale much shorter than a day, i.e. hours.

\section{Observations}\label{sec:obs}

\begin{figure*}[htb]
\centering
\includegraphics[width=1.0\textwidth]{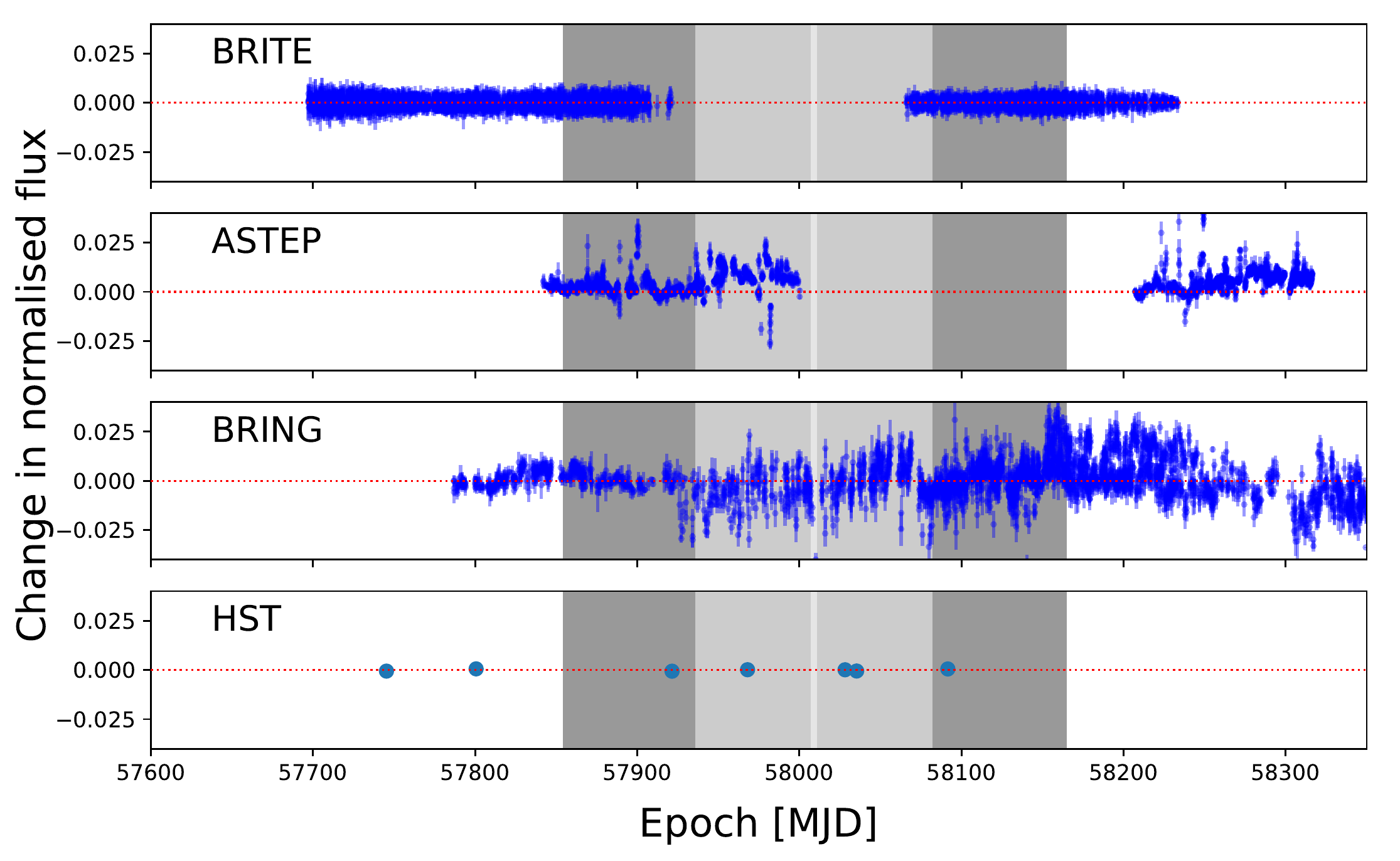}
\caption{Binned photometry of \bp{} for the four observatories. The transit of the Hill sphere of \bpb{} is shown as light grey and dark grey panels, representing the 100\% and 50\% radii of the Hill sphere, and the midpoint is the closest approach.}
\label{fig:binnedphot}
\end{figure*}

The reduction steps for each telescope are detailed in the Sections below.
To reduce the size of the photometric data sets we take a binned average of 0.05 days (72 minutes) for bRing, ASTEP and BRITE.
The photometric series from the four telescopes are shown in Figure~\ref{fig:binnedphot}.

\subsection{bRing and MASCARA}

To monitor the Hill sphere transit of \bpb{} for several months requires multiple dedicated observatories distributed in longitude.
To this end, the \bp{} Ring (bRing) observatories \citep{Stuik17} were built and deployed to Sutherland, South Africa and Siding Springs, Australia.
The first bRing observatory was built and tested at Leiden Observatory (PI: M. Kenworthy) and deployed almost exactly one year after the initiation of the project, with first light on 06 January 2017 at the South African Astronomical Observatory at Sutherland in South Africa - the details of the completely automated observatories are detailed in \citet{Stuik17}.
The second bRing observatory was built at the University of Rochester and deployed by S.~Mellon and E.~Mamajek to Siding Springs, Australia \citep{Mellon20Thesis}.
These observatories were based on the design and experience gained at Leiden Observatory with the MASCARA observatories \citep{talens2017}, aimed at accurate photometry of the brightest stars ($m_V\ < 8.4$).
The cameras do not have a filter in front of them, leading to an effective bandpass from 463nm to 639nm.
The bRing camera pixels are approximately 1 arcminute on a side, and the commercial photographic camera lenses used have a \ac{psf} that changes shape and size significantly across the field of view.
The cadence of bRing observations is one image every 12.8 seconds.
A custom pipeline \citep{talens2018} was written to take the bRing data and produce photometry with 1\% precision.
Although the two bRing observatories had almost complete longitudinal coverage for \bp{}, additional data was gathered from MASCARA-South, at La Silla Observatory, Chile, to enable redundant observations.
With a maximum observable zenith angle for the bRing stations of $\approx 80\deg$, \bp{} remained visible for at least 1 hour per night all year around.
During the Hill sphere transit itself, bRing took 9528 binned data points and each camera averaged 108 binned data points per night.

MASCARA and bRing are ground based observatories, observing using stationary, wide field cameras.
The data shows strong trends introduced by inter-pixel sensitivity variations, lens transmission, atmospheric transmission and weather, contamination by sun, moon light and neighboring stars.
For the calibration and detrending, a two step approach was used.
The initial calibration was performed according to the steps described in \citet{talens2018}.
This calibration performs a spatio-temporal calibration based on the average behaviour of all stars in the camera's field of view, over a baseline of approximately two weeks, and removes most of the spatial variation signatures in the point spread function and transmission, the variations in inter-pixel sensitivity, as well as variations in the atmospheric transmission due to clouds or dust. 
The residual systematic trends in the data vary from star to star, both on daily time scales as well as on monthly and yearly time scales, and are attributed to the sub-Nyquist sampling of the camera \ac{psf} by the lenslet array fixed on the interline readout CMOS array.
\citet{talens2018} describes several models for these individual trends in the data and their subsequent removal.
Here we use a modified approach of the ``local-linear'' method, where instead of fitting the sky background, we fit the moon phase and altitude and use them as an estimate for the sky background.
Similar to \cite{talens2018}, we iteratively solve with a 3-day moving mean to separate long term trends from the daily variability.

It is clear that there are residuals on the timescales of hours to days with amplitudes of up to 3\%.
When all three telescopes show photometric data, we see that sometimes the photometry from two of the three telescopes agree, with the third showing a deviation of up to 2\%.
We infer that these are due to systematics within that given telescope and not due to astrophysical phenomena associated with the \bp{} system.

\subsection{BRITE-Constellation}

The BRITE-Constellation\footnote{\url{https://brite-constellation.at/}} are a set of five nanosatellites each with a 3cm diameter telescope reimaging onto an uncooled CCD \citep{Weiss14}.
Three of the satellites -- i.e., BRITE-Toronto (BTr), Uni-BRITE (UBr) and BRITE-Heweliusz (BHr) -- observe with a red filter (550-700nm) and two -- i.e., BRITE-Austria (BAb) and BRITE-Lem (BLb) --  with a blue filter (390-460 nm). \citet{pablo2016} includes a detailed description of the detectors, pre-launch and in-orbit tests. 
The data is reduced with a custom pipeline \citep{Popowicz17} that processes the observed images and produces instrumental magnitudes that are delivered to the users.
The satellites observe a 24 square degrees wide field of view that contains 15-20 bright $(V<6)$ stars and at least three targets brighter than $V=3$ mag. Each field is observed for at least 15 minutes in each $\sim 100$ min orbit for up to half a year.
\bp{} was observed during three consecutive seasons with the BRITE-Constellation nanosatellites:
The first observations of \bp{} were obtained
from UT 2015 Mar 16 to 2015 June 2 (BRITE Run ID: 08-VelPic-I-
2015), yielding a total time base of 78.323 d using the BHr (red filter) satellite. 
A second observing run was conducted using BTr (red filter) from 
UT 2016 Nov 4 to 2017 Jun 17
for a total of 224.573 d and BLb (blue) from 
UT 2016 Dec 15 to 2017 Jun 21 for 187.923 d 
(BRITE Run ID: 23-VelPic- II-2016).

BHr was used from UT 2017 Jan 7 to 2017 Jan 30
for 24 days to cover a gap in the BTr observations.
During the third season, the red filter BHr satellite obtained time series of \bp{} between 
UT 2017 Nov 9 to 2018 Apr 25 for 167.335 days 
(BRITE Run ID: 33-VelPicIII-2017).
The BRITE-Constellation data of \bp{} are publicly available in the BRITE Public Data Archive\footnote{\url{https://brite.camk.edu.pl/pub/index.html}}.

In a next step, the raw photometric time series from the BRITE satellites were subsequently corrected for instrumental effects including outlier rejection, and both one- and two-dimensional decorrelations with all available parameters, in accordance with the procedure described by \citet{pigulski2018}.
A detailed description of the BRITE-Constellation data of \bp{} obtained during the three observing seasons and the corrections applied to them can be found in \citet{Zwintz19}.
In the present work, we used the same reduced and corrected light curves as in \citet{Zwintz19}. They are available on CDS\footnote{\href{http://vizier.u-strasbg.fr/viz-bin/VizieR?-source=J/A+A/627/A28}{http://vizier.u-strasbg.fr/viz-bin/VizieR?-source=J/A+A/627/A28}}.

\subsection{ASTEP}

Photometric observations were conducted with ASTEP, a 40\,cm telescope installed at the Concordia station, Dome C, Antarctica.
ASTEP is a Newtonian telescope equipped with a 5-lens Wynne coma corrector and a 4k $\times$ 4k front-illuminated FLI Proline KAF 16801E CCD with 16 bit dynamic range.
The corresponding field of view is $1^\circ\times 1^\circ$ with an angular resolution of $0.93"\rm\ pixel^{-1}$.
The effective bandwidth of the instrument and telescope is from 575nm to 760nm \citep{Abe13}.

At the latitude of Concordia, $75^{\circ}.01$S, $\beta$\,Pictoris is circumpolar, allowing a continuous monitoring during the antarctic winter season.
We observed it during two seasons, from 
2017 Mar 5 to 2017 Oct 14 and from 2018 Mar 5 to 2018 Jul 16.
Data acquisition started automatically when the Sun was 8$^{\circ}$ below the horizon, with a 30 sec exposure when the Sun was between 6$^{\circ}$ and 8$^{\circ}$ below the horizon (dawn and twilight), and 60\,sec otherwise.
Because of $\beta$ Pic's brightness, we used a Sloan $i'$ filter ($0.695-0.844\,\mu m$) combined with a highly defocused point spread function of about 100 pixels in diameter.
We performed aperture photometry on the images, retrieving lightcurves for Beta Pic and 17 comparison stars \citep[see][]{Mekarnia2017}.

The homogeneous set of lightcurves was in line with the excellent weather inferred from observations at Concordia between 2008 and 2012 \citep{Crouzet18}.
The $\delta$ Scuti variations are clearly visible in the day-to-day lightcurves \citep[][]{Mekarnia2017}.
The long-term stability of the lightcurves are however affected by two factors that were identified later.
First, the fact that $\beta$~Pic is about 13 times brighter than the first reference star implies that the correction for a varying background are less efficient than for usual observations of fainter stars \citep[e.g.,][]{Mekarnia16}.
Second, snow storms led to the deposition of ice crystals not only on the primary mirror, but also on the entrance window to the camera box, in a region where the optical rays are not parallel.
This led to global changes in the photometry of the target and reference stars, depending on where ice was deposited on the entrance window and on the location of the stars in the sky.
For MJD before 57970, HD~38891 ($\alpha$\,=\,05:46:11.9, $\delta$\,=\,-50:52:18; J2000) is used to calculate the daily median used to calibrate the data over one given night.
After a snow storm that introduced vignetting on MJD 57970, HD~38891 is used with a multiplicative factor of 0.985 up to MJD 57907.
A subsequent removal of ice crystals after 57907 changed the stability of the photometry, and HD~38745 ($\alpha$=05:45:11.7, $\delta$=-50:56:59)
 was used for calibration after this date.
Only photometry with the Sun more than 15 degrees below the local horizon is used ({\tt SUNELEV} $<-15^{\circ}$), data that is flagged as photometrically poor and any observations where the sky background rises above 200 counts are removed.

\subsection{Hubble Space Telescope}

Two \ac{hst} programs (GO-14621 and GO-15119; PI: Wang) obtained precision photometric data using WFC3/UVIS in spatial scanning mode.
In the first program, we monitored the flux of \bp{} over four visits within a span of 8.5 months (UT 22 Dec 2016, 16 Feb, 16 Jun, 2 Aug 2017).
%
The first two visits were timed to acquire a baseline (out-of-transit) constraint and the final two visits were timed to coincide with the predicted full Hill sphere ingress and transit.
We acquired two \ac{hst} orbits per visit in order to effectively model the star's variability.

The second program consists of three visits spanning just over two months (UT 1 Oct 2017, 8 Oct 2017, 4 Dec 2017).
%
These visits were timed to obtain precise constraints during half and full Hill sphere egress.
For this program, we only acquired one \ac{hst} orbit per visit because we found that the visit-to-visit variability during the first program was larger than the star's variability within a given visit.

We used WFC3's UVIS detector in 2K2C subarray mode (2k$\times$2k pixels, amplifier C) with the F953N narrow-band filter (953~nm).
For each frame, we scanned the star along the $x$ axis at a rate of 0.5\arcsec\,s$^{-1}$ 
for 110 seconds, thus spanning $\sim$1400 pixels per frame.
STScI recommends scanning in both forward and reverse directions to ensure that the target returns to the same point on the detector at each subsequent scan.
We performed at least five round-trip scans per \ac{hst} orbit; orbits with guide star reacquisitions permitted an additional forward scan.  All seven visits within both programs used this observing set up. 

\begin{figure*}
\includegraphics[width=1.0\textwidth]{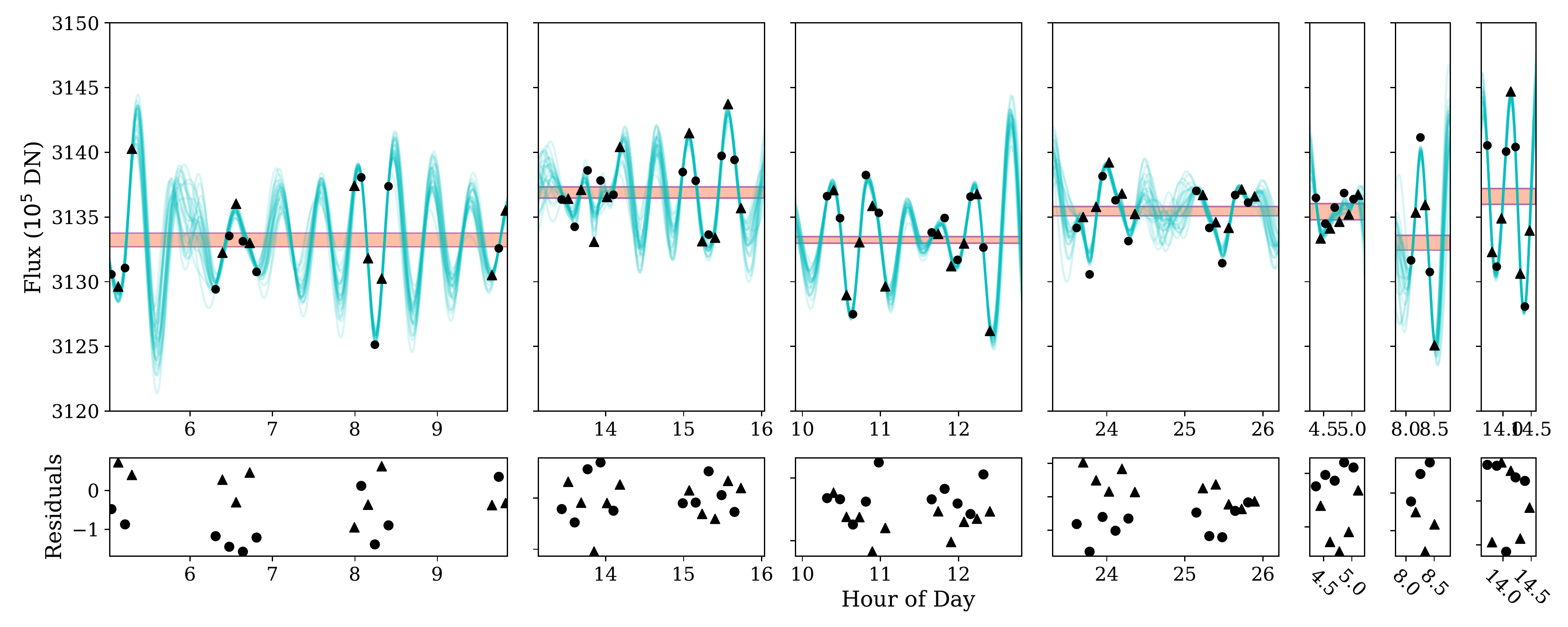}
\caption{\ac{hst}/WFC3/UVIS spatial scanning photometry obtained from all seven visits. The top panels show the data and model fits at each epoch. The forward scans are plotted as black circles, while the reverse scans are plotted with black triangles and have been scaled to correct the offset between the two scan directions. The teal lines are Gaussian process models of the photometry using Gaussian process parameters drawn randomly from our MCMC analysis. The horizontal red line represents the 1$\sigma$ statistical uncertainty on the flux level in each of the visits, and does not include the uncertainty of the photometry between visits. The bottom row show the average residuals to the fits.
\label{fig:visits}}
\end{figure*}

We use a custom pipeline to extract the time-series photometry from the WFC3-UVIS data.  Originally written for WFC3-IR analyses \citep[e.g.,][]{Stevenson2014a, Stevenson2014c}, the pipeline uses standard data reduction techniques that have been optimized for this particular observation.  For our final solution, we extract a 1500 $\times$ 700 pixel region centered on the scanned star, utilize a 1500 $\times$ 100 pixel rectangular aperture to determine the stellar flux, and use the remaining region for background subtraction.
The raw photometry from WFC3-UVIS contains a clear offset between the flux measured from the two spatial scanning directions.
We assume in this analysis that it differs by a multiplicative scale factor.
Given the known $\delta$ Scuti pulsation periods between 30--60 min \citep{koen2003a, koen2003b},
and the sparse time sampling but high precision of these \ac{hst} data, it is unfeasible to fit the over 30 known pulsation modes and not useful to use previous measurements of the pulsation that do not characterize the pulsations at sufficient precision \citep{Mekarnia2017}.

We are ultimately interested in the average flux for each visit.
Since each visit lasts multiple hours, we should sample over a full period of $\delta$ Scuti pulsations and thus retrieve the average flux value.
Adapting a similar approach as \citet{Johnson2015}, we model the stellar activity as a Gaussian process.
As we do not have sufficient cadence to sample the oscillations, we treat the $\delta$ Scuti pulsations as a quasiperiodic Gaussian process, where the periodic term roughly describes the strongest pulsation mode, and the ``quasi'' term accounts for the fact the modes constructively and destructively interfere, causing the amplitude to change in time.
We parameterize the quasiperiodic kernel as the product of a Mat\'ern kernel and a periodic kernel:

\begin{equation}
 K_{ij} = A^2 \cos\left(2\pi t_{ij}/P_{osc}\right) (1 + \sqrt{3}t_{ij}/l) \exp\left(-\sqrt{3}t_{ij}/l\right).
\end{equation}

Here, indices $i$ and $j$ refer to two data points separated in time by $t_{ij}$.
We assume times between visits are so far apart that there is no correlation, but we assume they are drawn from the same Gaussian process.
$P_{osc}$ is roughly the period of the dominate pulsation mode, and $l$ is the covariance length that damps correlation at long time baselines, making the kernel quasiperiodic. 

We assume that all seven epochs have $\delta$ Scuti pulsations that can be modeled by the same Gaussian process and that the flux offset between the two scan directions is the same multiplicative factor.
Running the following analysis on each individual visit did not indicate that any of these parameters were different.
We ran a Bayesian parameter estimation to fit for the flux in all seven epochs as well as the three Gaussian process parameters ($A$, $P_{osc}$, $l$) and the multiplicative favor to correct for the offset between the scan directions.
We also fit for a term that increases the error of each flux measurement above the nominal photon noise term to account for unknown effects such as the imperfection of the Gaussian process kernel in fully modeling the observed stellar activity.
We use uniform priors on the flux at each epoch, and log-uniform priors on all of the nuisance parameters.
We used \texttt{emcee} \citep{ForemanMackey13} to sample the posterior of fluxes, marginalizing over all nuisance parameters.
The Gaussian process regression was implemented with our own custom code.

Looking at the nuisance parameters, we find a multiplicative scale factor of $0.9984 \pm 0.0001$ for the fluxes from the second scan direction.
Our Gaussian process finds a period of $29.3 \pm 0.8$~minutes, which is just smaller than all of the known pulsation periods.
This shorter period might allow the Gaussian process model to best fit the full range of pulsation frequencies.
We found that we needed to scale the uncertainties on the photometry of each scan by $1.9 \pm 0.1$ times the photon noise limit to account for the scatter in our measurements.
It is difficult to determine whether the additional noise above the photon noise floor is due to the instrument, data reduction, or the Gaussian process not being a perfect model of the pulsations. 
The HST data and the Gaussian process model used to measure the average flux in each epoch is shown in Figure~\ref{fig:visits}.

The statistical uncertainty on the flux in each epoch from our analysis is 0.01--0.02\%.
However, comparing the two out of transit observations, we find a difference of 0.11\%, which is likely limited by the stability of the telescope between visits. 
This is consistent with a finding of 0.1\%  repeatibility for UVIS drift scans by the instrument team \citep[Instrument Science Report WFC3 2017-21;][]{Shanahan2019}.
It is therefore reasonable to expect a similar amplitude of uncertainty during all in-transit visits, and we use 0.06\% as our 1$\sigma$ uncertainty in the flux of each epoch.
Unlike typical exoplanet transit observations with \ac{hst} where the out of transit and in transit photometry can be obtained in the same visit, we are relying on the photometric stability from visit to visit. 

Due to the uniqueness of the transit, we would have followed up on anything with greater than a 3$\sigma$ deviation from the out of transit flux.
We did not observe the star to dim significantly in any of our visits, so we establish the 3$\sigma$ value of 0.18\% as our sensitivity limit.

\subsection{Summary of photometric data}

The photometric measurements from HST suggest that there are no significant variations in the photometry of Beta Pictoris at the times of the seven visits.
In each of the two separate seasons of BRITE photometry, no long term variation is seen.
HST shows that there is no relative offset between the BRITE seasons, implying that there is no long-term astrophysical variation during the observations from the space based observatories.
%
%
HST and BRITE therefore provide a check of the variations in the photometry seen in the BRING and ASTEP data during the Hill sphere transit, where we see larger variations.
We therefore hypothesise that any possible astrophysical fluctuations in the ground-based photometry are below the level introduced by time-varying systematic effects, which are on the order of 2\%.

\section{Analysis}\label{sec:analysis}

We search for occulting material within the Hill sphere of \bpb{} by looking for long term photometric changes as a function of the Hill sphere radius, with timescales of weeks to months.
Due to the difficulties of removing long term systematics (and the danger of possibly removing any possible astrophysical signal) we hypothesise that there is no statistically significant \ac{cpd} detection with the BRITE photometry and that for parameter values outside of BRITE's coverage, we can use the ground based observatories to provide upper limits on $\tau$.
For this analysis we analyse the photometry from each telescope independently and then combine the results into a final sensitivity plot.

\subsection{Dust properties}\label{dust}

Our model derives limits on occulting material in terms of optical depth, and to convert this to estimated limits on dust mass we make some simple assumptions.
Using Equation \ref{chen1} we solve for the temperature of a dust grain ($T_{g}$) at a given distance $D$ from a source of temperature $T$ (in K) and radius $R$ \citep{Chen01}:

\begin{equation}
\label{chen1}
T_{g} = \sqrt{\frac{R}{2D}}T
\end{equation}
Using values from Table~\ref{bpicparams} yields an equilibrium temperature of $\sim$174 K due to stellar radiation, and an equilibrium temperature of $\sim$101 K from the planet's thermal emission.
Even at a distance of 8.9\,au, the star's flux dominates that of the planet (the planet provides negligible heating beyond 0.01$r_H$). Ices will likely sublimate at these temperatures, so we adopt silicate as the dominant dust grain composition and adopt a density of $\rho_g$ = 2.5 g cm$^{-3}$, corresponding to the density measured by \citet{Chen01} in Jupiter's rings.

Given the age of $\beta$~Pictoris, and the fact that the star itself has largely dispersed its primordial disk material, it is unlikely that any primordial gas-rich circumplanetary disk survives.
We therefore assume that any circumplanetary dust is replenished through collisions of larger objects, so can be thought of (and modelled) as a microcosm of a circumstellar debris disk \citep[e.g.][]{Kennedy11}.
Typically, collisional dust size distributions are such that most of the surface area is concentrated in the smallest surviving grains \citep{1969JGR....74.2531D}.
Thus, we estimate the minimum grain size for circumplanetary orbits using equation (9) of \citet{Kennedy11}; while this minimum size is analogous to the radiation pressure ``blowout'' size for circumstellar orbits, the smallest circumplanetary grains may also collide with the planet (or any moons) as their orbits are driven to high eccentricity \citep[see][]{Burns79}.
The true minimum grain size depends on the specific orbit, but this estimate is sufficient for our purposes here.
Assuming that dust is concentrated at the area-weighted mean planetocentric distance (0.7$r_{CPD}$, see below), the minimum size is $s = 31 \sqrt{r_{CPD}/r_{Hill}}$\,$\mu$m, approximately six times larger than the blowout size for circumstellar orbits.

\subsection{Circumplanetary disk model}\label{mellonestimate}

The rings of the gas giant planets in the Solar system are perpendicular to the rotational axis of the parent planets, marshalled there by the quadrupole moments of the planet's gravitational field.
At larger radii from a planet, it is expected that the rings would become coplanar with the planet's orbit, and \citet{Speedie20} investigate the stability and extent of tilted ring systems around exoplanets. 
A determination of the rotational period of \bpb{} (for example by photometric monitoring) together with the radius of the planet would enable an estimate of the planetary obliquity projected onto the line of sight towards Earth.
The obliquity of \bpb{} is not known, although a measurement of rotational broadening by \citet{Snellen14} implies that the planet is not being viewed pole on.
Given that the four gas giant planets in the Solar system have a range of obliquities from 3 degrees to 98 degrees, it is reasonable to assume that the obliquity of \bpb{} is unconstrained, and that the angle between the rotational axis of the planet and its orbital plane is similarly unconstrained, although it is worth noting that the spin axis of the star and the orbit of \bpb{} are coaligned within measurement errors \citep{Kraus20}.
One might argue that any \ac{cpd} would be coplanar with the planet's orbital plane, but simulations by \citep{Martin20} show that \ac{cpd}s with small initial tilts can have a tilt instability increase its tilt and possible move them into our range of detection. 
Once tilted, the stability of inclined \ac{cpd}s from \citet{Speedie20} show that they can last on long timescales and remain detectable in transit.
Coplanar \ac{cpd}'s are truncated at 0.4\rhill{} \citep{Martin11} but a tilted \ac{cpd} may have a larger truncation radius \citep{Lubow15,Miranda15}.
For the purposes of this analysis, we assume that a \ac{cpd} can be allowed at any orientation.

We construct a simple model of the eclipse light curve for a circumplanetary disk.
We assume that the disk has a height much smaller than its diameter, so we can approximate it as a thin slab of homogeneous material with face-on optical depth $\tau$.
We use a coordinate system whose origin is the centre of the disk, with the positive z axis pointing towards the observer.
The disk is circular with a radius $r_{CPD}$ centered on the origin and lies initially in the xz plane.
It is inclined by rotating about the x axis by $\theta$ degrees, and is then rotated around the z axis by $\phi$ degrees in the direction from the positive x axis towards the positive y axis, as shown in Figure~\ref{cpdmodel}.

The unattenuated star flux has a value of $I_0$.
The star light passing through the disk towards the observer is then attenuated as:

$$I=I_0 \exp (-\tau\sin \theta)$$

We assume that $\tau << 1$ and we Taylor expand to give:

$$I = I_0(1-\tau\sin \theta)$$

and rearrange to get:

$$\tau = \frac{(1-I/I_0)}{\sin \theta} $$

The surface density of the disk $\sigma_{CPD}$ is given by $\tau/\kappa$, and the total mass of dust in this disk is then:

$$M_{CPD}=\frac{\tau}{\kappa} \pi r_{CPD}^2$$

where $\kappa$ is the opacity in units of cm$^2$ g$^{-1}$ and can be written in terms of dust density $\rho$ and particle size $a$ as $\kappa=3/(4\rho a)$, leading to:

$$M_{CPD}= \frac{4\tau a \rho}{3} \pi r_{CPD}^2$$

The x-axis is parallel to the projected path of the star behind the disk, the y-axis is oriented such that the path of the star crosses the y-axis at impact parameter $b$ when $t = t_{b}$ at $x=0$, so that the coordinates of the star at time $t_b$ is $(0,b)$ and the x coordinate at time $t$ is:

$$x_{star} = v(t-t_{b})$$

In this way, we can calculate a model light curve $I(t)=f(r_{CPD},\tau,i,\phi,t_b,t)$.

\begin{figure}[htb]
    \includegraphics[width=\columnwidth]{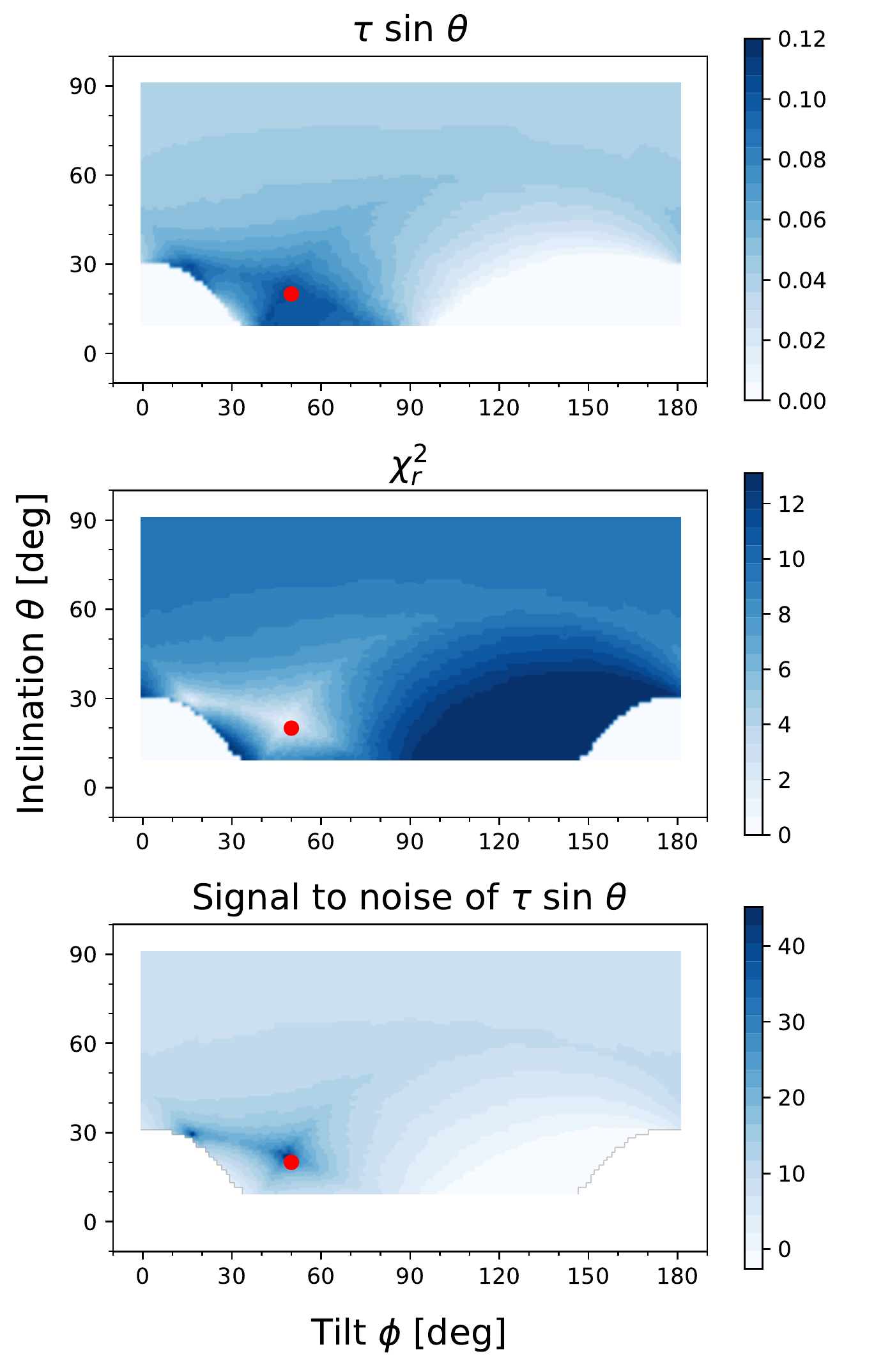}
    \caption{Fitting to a synthetic \ac{cpd} dataset showing the estimated $\tau$, reduced chi squared, and the signal to noise of the measurement of $\tau$.
    The red dot indicates the input inclination and tilt of the best fitting disk.}
    \label{simdisk}
\end{figure}

For each instrument, we have the photometric time series $I(t)$ and the error on the measured flux $I_{err}(t)$.
We fix the radius of the disk $r_{CPD}$ and generate a grid of trial values for the orientation of the disk in $(i,\phi)$.
With each pair of trial values, we calculate the reduced chi squared of the model with respect to the data, and we use the Python module {\tt lmfit} to perform the minimisation and find the best fit $\tau$ value for the model.
An example disk and data set is presented in Figure~\ref{simdisk} for a disk of radius 0.40\rhill{}, a best fit optical depth of 0.1, at an orientation $\theta=20^o, \phi=50^o$.
Contours of higher values of $\chi_{r}^2$ are not symmetric about the best fit but show narrow regions corresponding to disk geometries where the chord cut across the disk is of a similar length to the chord of the disk with the best fit.

\begin{figure*}[htb]       
    \includegraphics[width=0.333\textwidth]{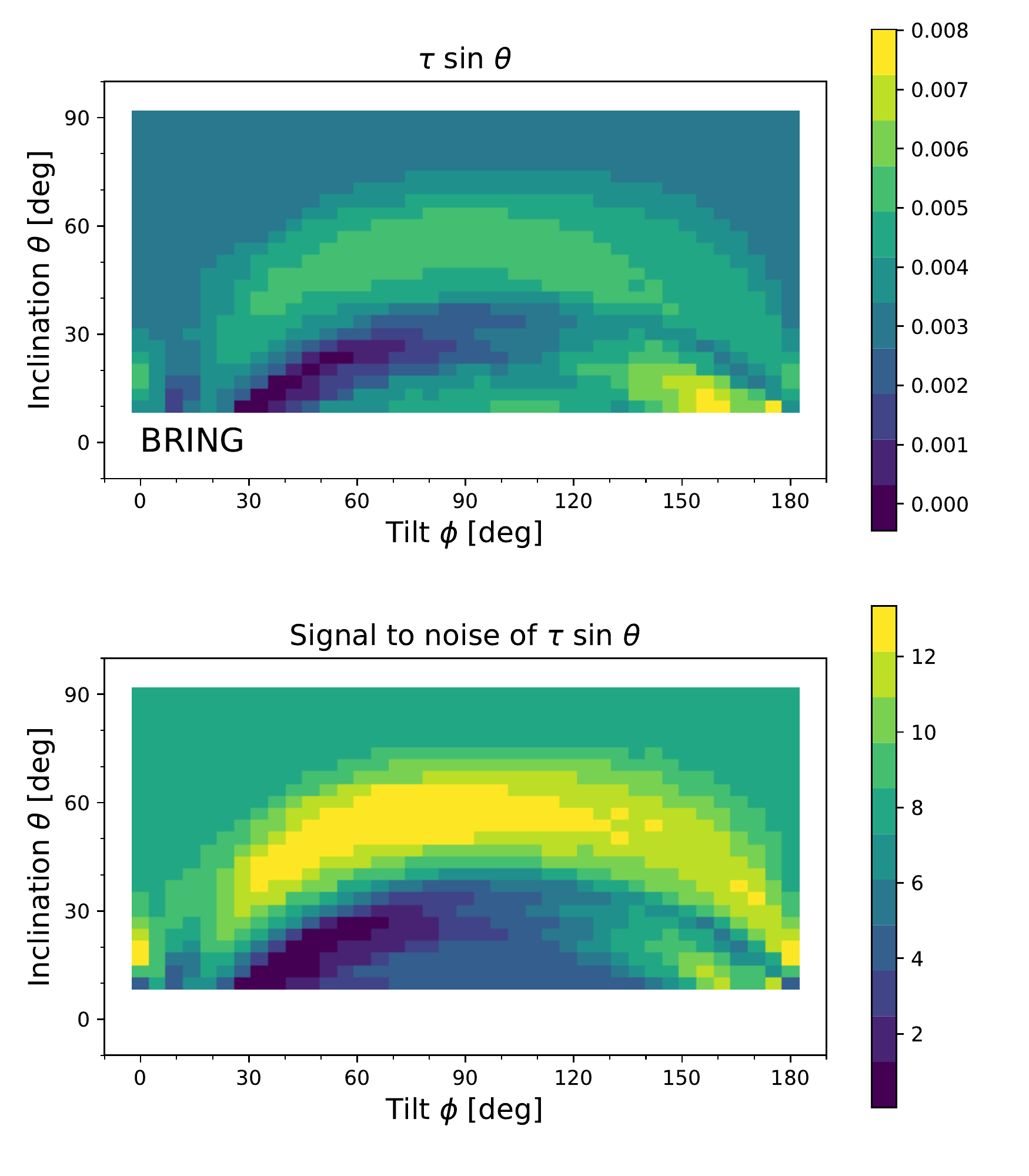}   
    \includegraphics[width=0.333\textwidth]{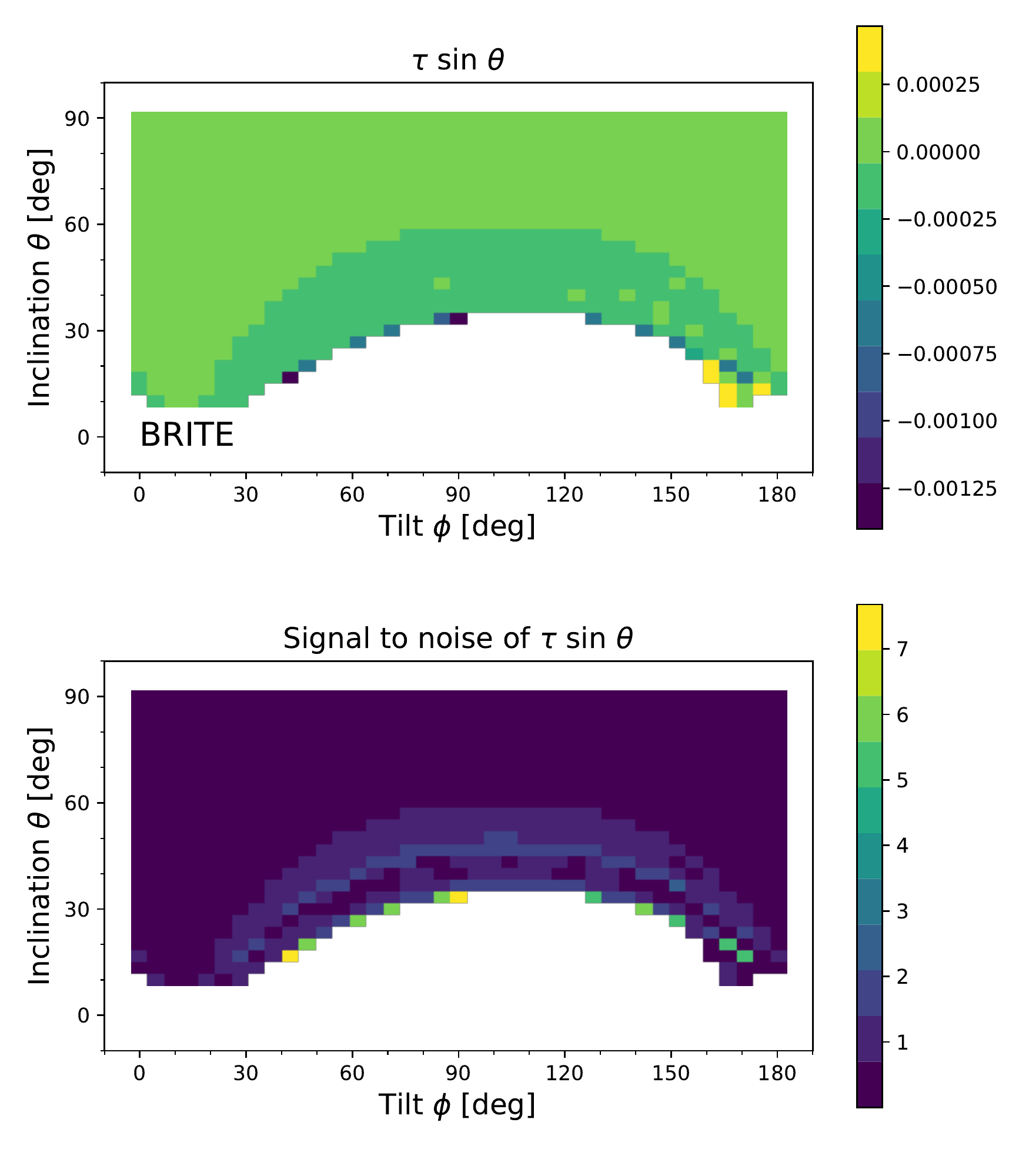}
    \includegraphics[width=0.333\textwidth]{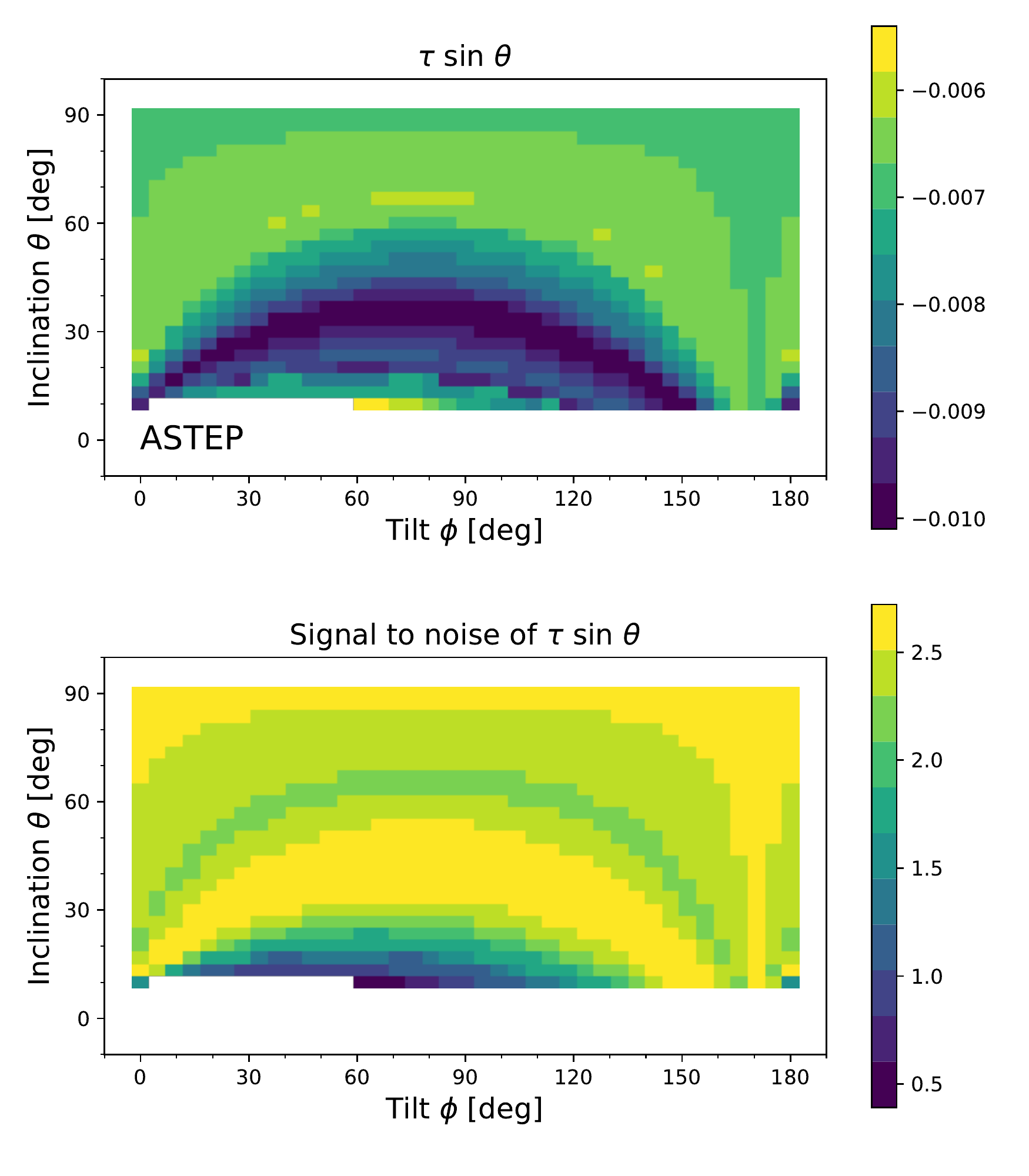}
    \caption{Circumplanetary disk fits for a disk with radius 0.60\ \rhill. The three observatories are shown in the three panels. The top row shows the upper limit on the optical depth for a given tilt and inclination of the model disk. The numerical values of the upper limits are shown in the colour bar on the right of the plots. White areas are where there is no photometry with the given observatory to provide a constraint. The lower row shows the signal to noise for each trial value of tilt and inclination.}
    \label{cpd60}
\end{figure*}

\begin{figure*}[htb]       
    \includegraphics[width=0.33\textwidth]{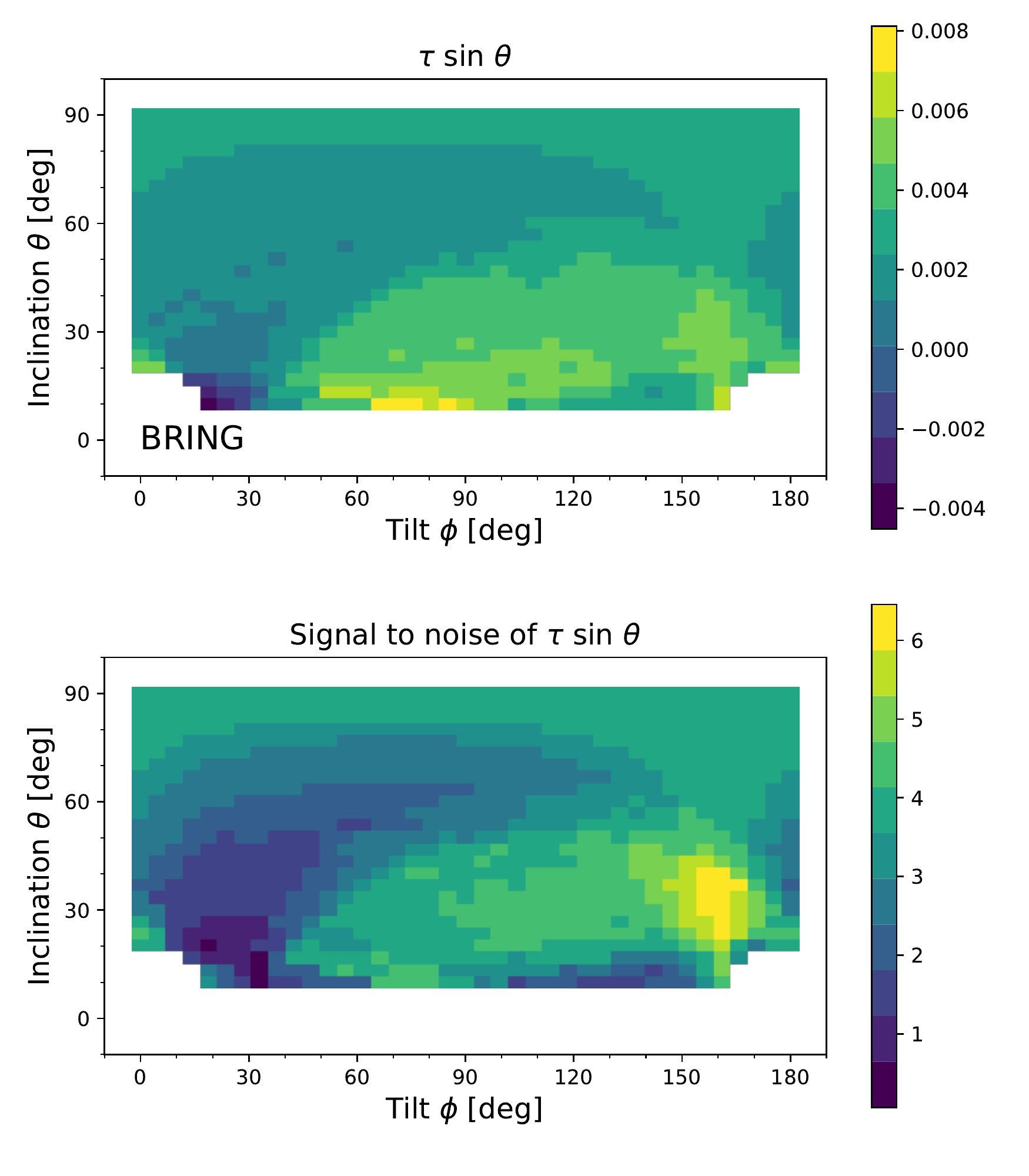}   
    \includegraphics[width=0.33\textwidth]{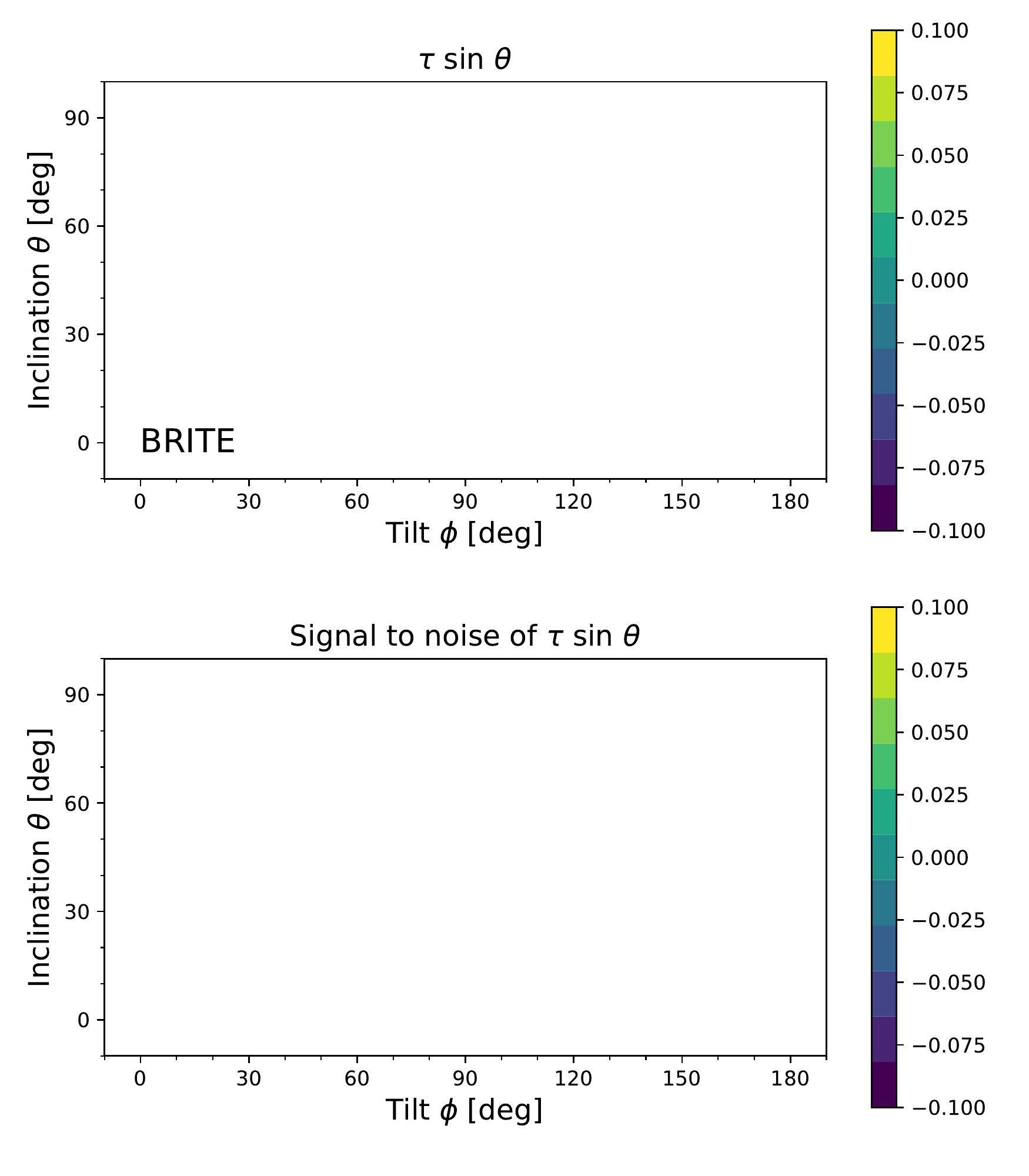}
    \includegraphics[width=0.33\textwidth]{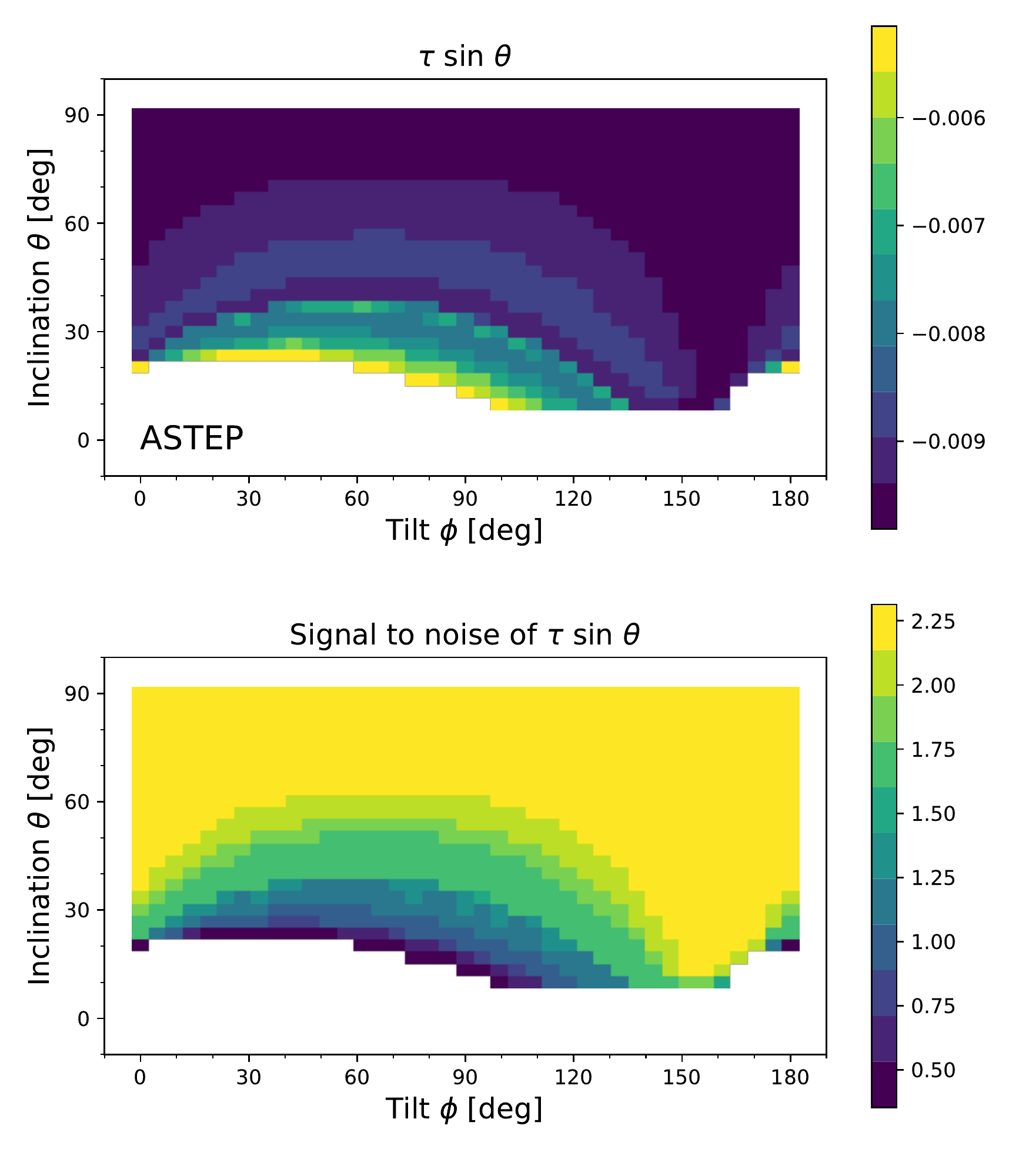}
    \caption{Circumplanetary disk fits for a disk with radius 0.30\ \rhill. The three observatories are shown in the three panels in a similar format to Figure \ref{cpd60}. The smaller radius for the CPD leads to different amounts of coverage within the tip and tilt parameter space. BRITE does not have photometric coverage to test any CPD model with a radius of 0.30\ \rhill.}
    \label{cpd30}
\end{figure*}

We produce maps of fitted optical depth $\tau$ for each of the three instruments, and for two disk radii, 0.6\rhill{} (Figure~\ref{cpd60}) and 0.30\rhill{} (Figure~\ref{cpd30}).
Each of the observatories has a different temporal coverage of the transit, and so they probe different regions of parameter space for possible \ac{cpd} orientations.
The BRITE photometry shows no significant photometric systematics, whilst the two ground based observatories bRing and ASTEP show significant non-zero values for $\tau$, represented as positive values of $\tau / \tau_\sigma > 1$ in the lower panels. 
For a 60\% Hill sphere \ac{cpd}, we see that BRITE provides the most sensitive upper limits on the optical depth, but that bRing provides the most complete coverage of possible tilts and inclination.
For the 30\% Hill sphere \ac{cpd} the BRITE satellite coverage does not put any constraints on any possible \ac{cpd} (see the light gray regions in Figure~\ref{fig:binnedphot}).
The almost continuous coverage from bRing provides complete photometric coverage for smaller \ac{cpd}s, at a cost in precision.

The long term photometric monitoring places an upper limit on the mass of a \ac{cpd} around \bpb{} for geometries where a disk would intersect the chord drawn by the star behind the Hill sphere.
For a stable prograde (0.3\ \rhill{}) and retrograde (0.6\ \rhill{}) \ac{cpd}, the mass limits are $2.2\times 10^{21}$g and $1.8\times 10^{22}$g respectively. 
%

\begin{figure*}[htb]       
    \includegraphics[width=0.50\textwidth]{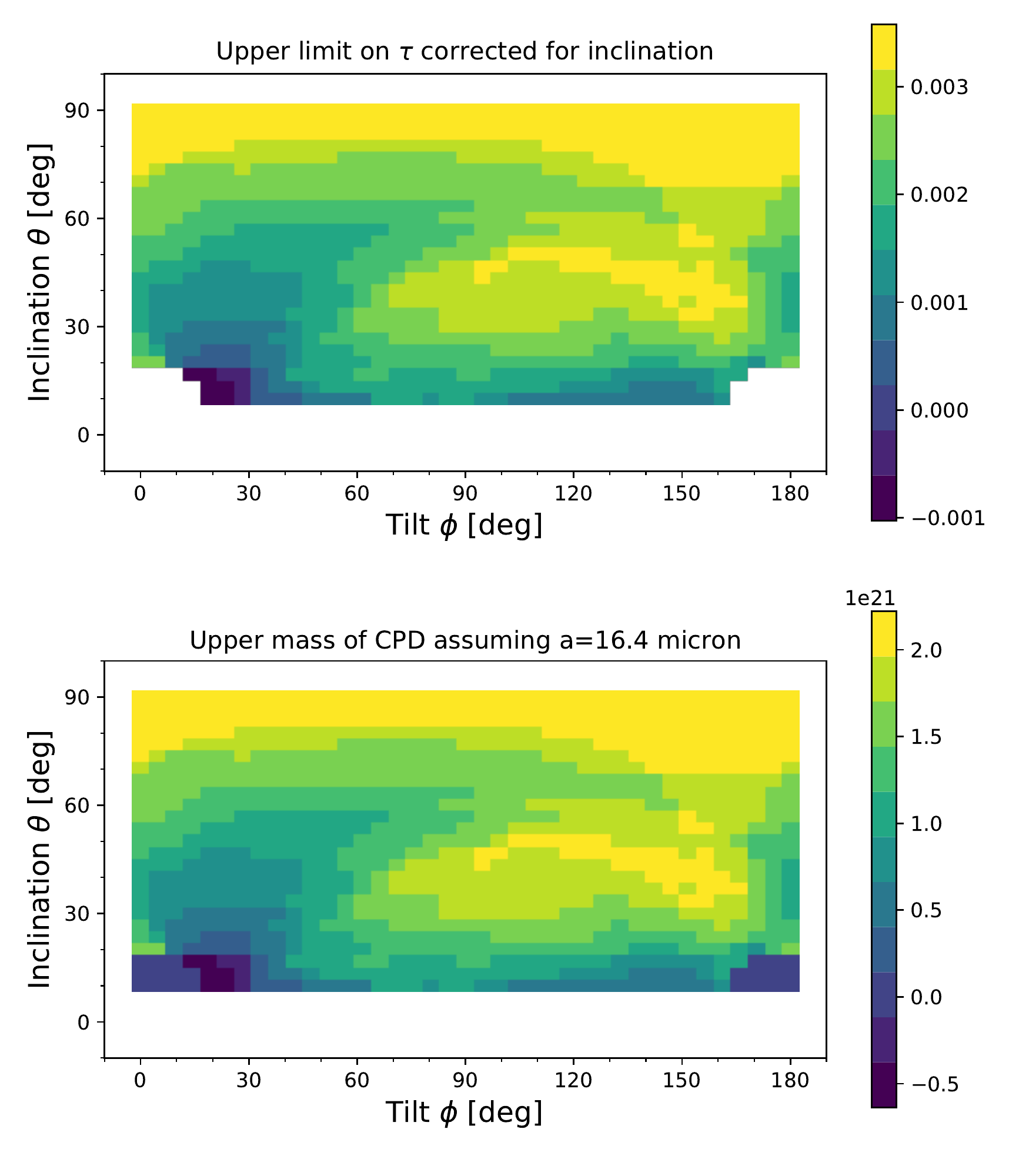}   
    \includegraphics[width=0.50\textwidth]{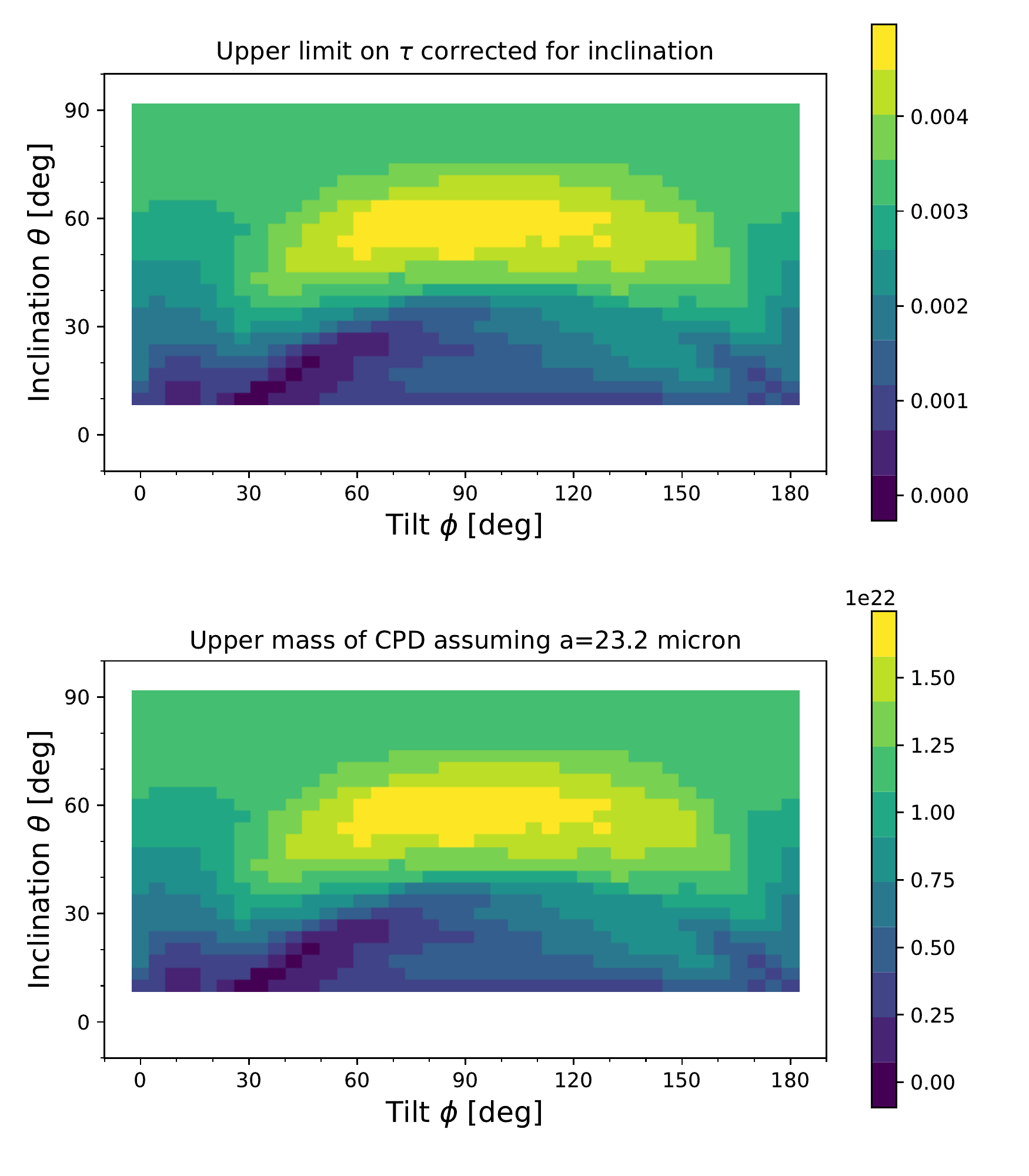}
    \caption{\ac{cpd} models for 0.30\rhill{} and  0.60\rhill{} radii. The upper row shows the $tau$ corrected for disk inclination, and the lower panel shows the upper limit on the total mass of the disk assuming mean particle sizes of $16.4\mu m$ and $23.2\mu m$.}
    \label{totalcpd}
\end{figure*}


\subsection{The 1981 Event}

A significant photometric fluctuation was seen towards \bp{} around 10 November 1981, and subsequently reported in \citet{LecavelierdesEtangs95}.
This variation was seen in five separate colour filters in the optical bands, and appeared to have no significant colour component \citep{Lamers97}.
Plausible models that could explain these photometric fluctuations include a horseshoe shaped cloud of dust following an orbit of a (then hypothesised) gas giant planet \citep{LecavelierdesEtangs97} or the large tail of an evaporating falling body with a comet-like tail structure \citep{Lamers97}.
One possible explanation for the 1981 event is that the event was generated by constructive interference of the $\delta$ Scuti pulsations from \bp{} itself, but the observed pulsations of \bp{} can only constructively increase the total flux by $\sim$0.1\% over timescales of a few minutes, and so we rule this out as an explanation for the 1981 event.
Another explanation is that the event is a result of systematic error in the original observations, but since the effect was observed simultaneously in several optical bands, this is considered highly unlikely \citet{LecavelierdesEtangs95}. 
Although the photometry is relatively sparse, one simple model is forward scattering at small angles from a cloud of small particles that do not directly block the disk of the star, combined with a much shorter duration transit event that brings the flux back to the nominal level of starlight.

\begin{figure}[htb]
\centering
\includegraphics[width=\columnwidth]{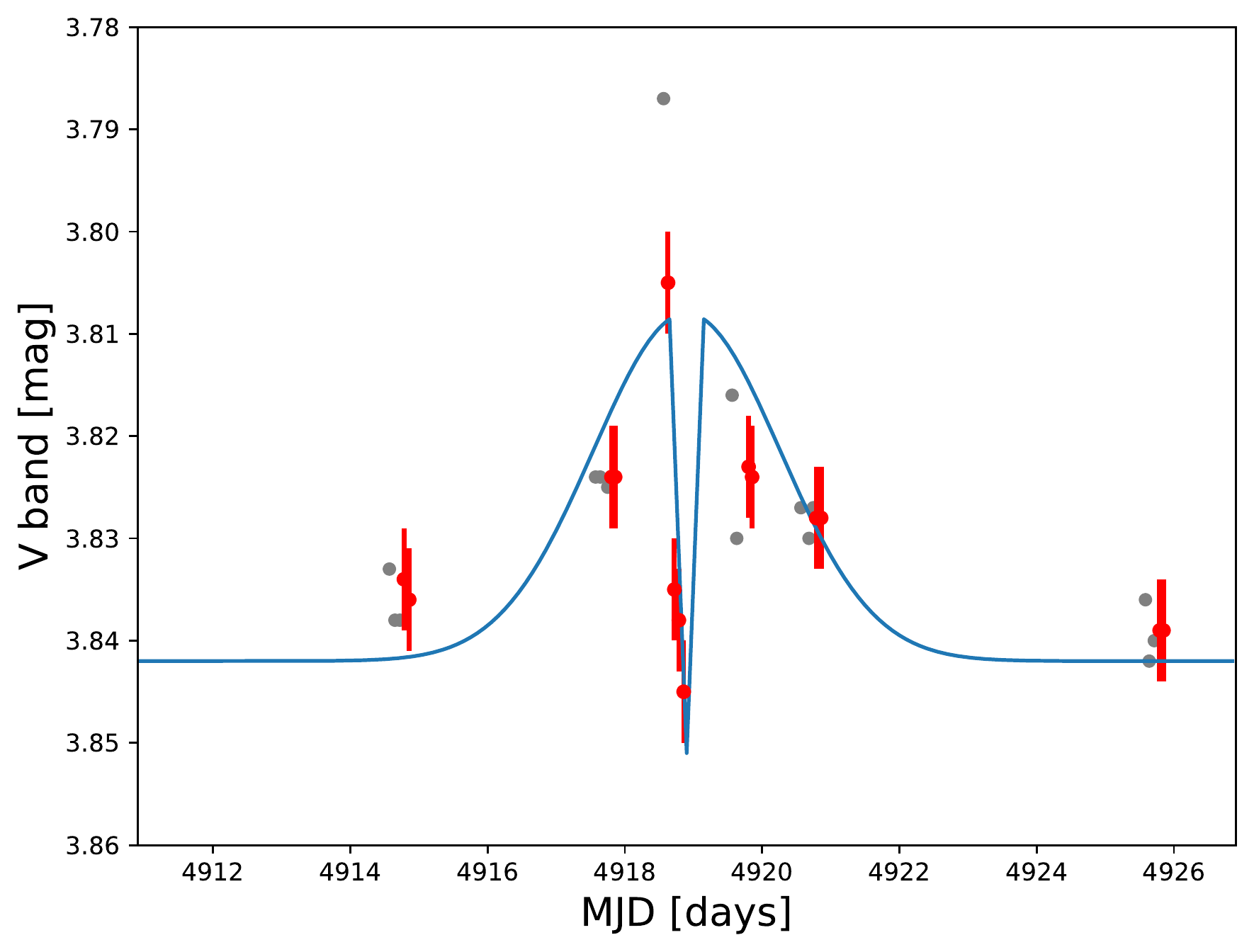}
\caption{The model for the 1981 event. The grey photometric points are V band magnitudes reported in \citet{LecavelierdesEtangs95}, and the red points are the subset of the photometry that passed quality checks from that paper. The red error bars were reported in \citet{Lamers97}, and the parameters for the model (shown in blue) were derived from those reported in \citet{Lamers97} Figure 7.}
\label{fig:1981model}
\end{figure}

The light curve is shown in Figure~\ref{fig:1981model} with a simple forward scattering model fitted to the measured photometry as described in \citet{Lamers97}.
An astrometric fit by \citet{Wang16} showed that the planet does not transit the star and so is not responsible for the 1981 event.
Instead we consider if a circumplanetary ring is responsible for the 1981 event, where an optically thick and narrow ring sits within a broader optically thin ring.
This ring is centered on \bpb{} and since the radius of the ring is much larger than the diameter of the star, the segment of the circumplanetary ring that crosses the stellar disk can be approximated by a straight line.
We follow \citet{Lamers97} and model the 1981 event as a Gaussian with FWHM of 3.2 days and amplitude of 0.035 magnitudes to model forward scattering from the optically thin part of the ring, and a notch feature representing the optically thick part of the ring, shown in Figure~\ref{fig:1981model}, such that the model $F_{1981}(t)$ has the midpoint of the model at $t=0$.
We then fit the photometry of \bp{} to see if we see a similar ring transit feature during the transit.

\begin{figure*}[htb]
\centering
\includegraphics[width=0.9\linewidth]{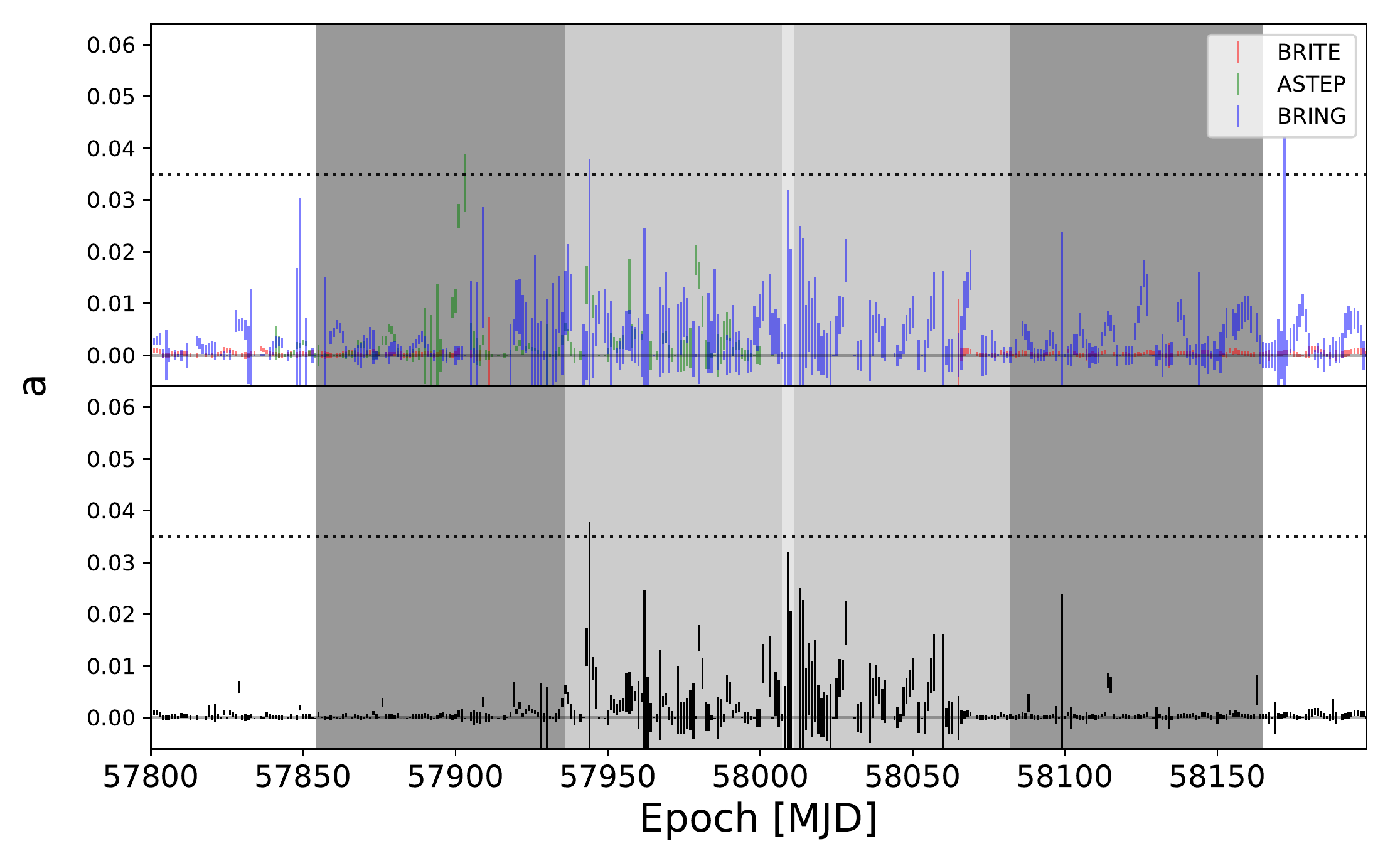}
\caption{Result of fitting the 1981 model light curve (shown in Figure~\ref{fig:1981model}) to the data from BRITE, ASTEP and bRing. The upper panel shows the value of the fit amplitude $a$, and the dotted line shows the measured amplitude from the 1981 event. The error bars are one sigma limits determined from the {\tt lmfit} algorithm. The lower panel shows the smallest value of $a$ if there is more than one telescope with data. The dark grey and light grey panels indicate the 100\% and 50\% radii of the Hill sphere.}
\label{fig:1981fit}
\end{figure*}

For each telescope, we choose a test epoch $t_{mid}$ for the midpoint of the 1981 model, and then fit a two parameter model:

$$F(t) = a F_{1981}(t_{mid}) + b$$

where $a$ and $b$ are free parameters.
The parameter $a$ is the amplitude of the model and $b$ is a constant offset in flux that accounts for any long term trends in the photometry from the star and telescope systematics.
A nonlinear fitting routine then takes a trial value of $t_{mid}$ and returns the best fit values of $a$ and $b$ for each test epoch.
The parameter $t_{mid}$ is chosen from MJD 58700 to 58200 in steps of 1 day.
The routine returns $a$ and its error, and the results for all three telescopes are shown in Figure~\ref{fig:1981fit} in the upper panel.
As for the CPD fitting, each telescope is treated separately, represented by the different colour points and error bars on $a$.
The dotted line represents an amplitude $a$ consistent with the 1981 event.
The trial time $t_{mid}$ is ruled out if the amplitude $a$ and its error bar does not reach the 1981 event level at $a=0.035$.

Due to the systematic errors in the ground based observatories, the fitted parameters do not always agree with each other within their respective error bars.
Under the hypothesis that there is no 1981 transit event in the Hill sphere of \bpb{} we take the lower of the two determined values of $a$, resulting in the lower panel of Figure~\ref{fig:1981fit}.
There are a few trial days where no model fit is determined due to lack of data, with the longest gap being around 57920.
The fits on days adjacent to these gaps are consistent with no 1981 event, so we consider it highly unlikely that a transit event was missed.
On MJD 57942 the error bar from bRing is larger than 0.04, meaning that it does not rule out a 1981 transit event, but the two days either side of this event are significantly inconsistent with a 1981 event.
We therefore conclude that there was no 1981 transit-like event throughout the whole passage of the Hill sphere of Beta Pictoris b.

\section{Discussion and Conclusions}\label{sec:concl}

This paper presents the first photometric monitoring campaign of the Hill sphere of a gas giant exoplanet beyond the ice line of the host star.
Several observing campaigns led by different research groups have combined their data to provide continuous coverage of the 300 day duration of the transit, making this analysis possible.
We detect no signal consistent with a \ac{cpd} crossing the line of sight within the Hill sphere placing an upper limit of $1.8\times 10^{22}g$ on any possible \ac{cpd} under our detection limits.
There are several interpretations to our results:
\begin{enumerate}
    \item There is a dust disk that has a projected radius smaller than 10\% of the Hill sphere radius, so it does not transit the star.
    \item There is a larger dust disk that has a low obliquity and does not transit the star.
    \item There are no significant amounts $(<1.8\times 10^{22}g)$ of circumplanetary micron sized dust in the Hill sphere.
\end{enumerate}

The coplanarity of moons around the gas giants in our Solar system implies the existence of circumplanetary disks at earlier epochs, and given the large amount of dust in \bp{} system, it is almost certain that there must have been a \ac{cpd} around \bpb{}.
We conclude that the circumplanetary dust has already condensed into moons, or is in the form of a disk with a projected radius smaller than 10\% of the Hill radius, or that there is a CPD below the sensitivity limit of our observations.
This is corroborated by ALMA observations that place upper limits on all the millimetre-sized dust in the Hill sphere of the planet \citep{Perez192}.
Our upper mass limit is over 60 times smaller than the limit placed by ALMA observations.

A transit event in 1981 was hypothesised to be due to either a circumstellar clump of material, an exocomet tail, or a narrow ring associated with a planet.
Our temporal coverage and photometric precision rules out a similar transit event, and so we conclude that the 1981 event was not due to dust in the circumplanetary environment of \bpb{}.
Whether the 1981 event was a singular, transitory event within the \bp{} system, or is a long lived structure associated with one of the planets within the system, remains to be seen.
The 1981 transit event remains unexplained.
We are continuing to monitor \bp{} with the bRing stations in South Africa and Australia, to see if we detect any material at the L3 and L4 Lagrange points in 2022.
The announcement of a second planet, \bp{} c, with a 6 year orbital period, brings up the possibility of a second Hill sphere transit that we can monitor for rings or other circumplanetary material.

It is reasonable to assume that there are other Hill sphere transits around young stars, and we are continuing our searches in both archival data and in Evryscope and TESS data, using Gaia DR2 proper motions to identify PMS stars ($<20Myr$) for candidates.
The ultimate goal is to identify a Hill sphere transit from archival photometric observations, to plan a high time cadence spectrophotometric campaign where the circumplanetary environment of a forming gas giant planet and its attendant moons can be studied, and determine their physical and chemical composition.
Hill sphere transits remain an exciting prospect for studying spatial and spectral scales of circumplanetary environments that are not possible with other imaging techniques.

\section{Source Code}
We are committed to open science, and have made the data, reduction scripts and plots in this paper available open-source.
They available at \url{https://github.com/mkenworthy/beta_pic_b_hill_sphere_transit}.
All code is provided under a BSD 2-Clause ``Simplified'' license. 


\begin{acknowledgements}

  We thank the referee for taking the time to review our paper, especially after the difficulties and disruptions of the past year.
  MAK acknowledges funding from NOVA and Leiden Observatory for the bRing observatory at SAAO, and to the NSF/NWO for travel funding (NWO grant 629.003.025).
  JW is supported by the 51 Pegasi b Fellowship.
GMK is supported by the Royal Society as a Royal Society University Research Fellow.
  MAK thanks the staff and observatory support crews at the South African Astronomical Observatory in Sutherland for all the work they put in to make bRing a successful observing station, and which allowed us to obtain first light within the first week of installation.
  Part of this research was carried out at the Jet Propulsion Laboratory, California Institute of Technology, under a contract with the National Aeronautics and Space Administration (80NM0018D0004).
Construction of the bRing observatory sited at Siding Springs, Australia was made possible with a University of Rochester University Research Award, help from Mike Culver and Rich Sarkis (UR), and generous donations of time, services, and materials from Joe and Debbie Bonvissuto of Freight Expediters, Michael Akkaoui and his team at Tanury Industries, Robert Harris and Michael Fay at BCI, Koch Division, Mark Paup, Dave Mellon, and Ray Miller and the Zippo Tool Room.
The results reported herein benefitted from collaborations and/or information exchange within NASA’s Nexus for Exoplanet System Science (NExSS) research coordination network sponsored by NASA’s Science Mission Directorate.
  This research is based on observations made with the NASA/ESA Hubble Space Telescope obtained from the Space Telescope Science Institute, which is operated by the Association of Universities for Research in Astronomy, Inc., under NASA contract NAS 5–26555. These observations are associated with programs 14621 and 15119.
ASTEP benefited from the support of the French and Italian polar agencies IPEV and PNRA in the framework of the Concordia station program and of Idex UCAJEDI (ANR-15-IDEX-01).
This research made use of Astropy,\footnote{http://www.astropy.org} a community-developed core Python package for Astronomy \citep{astropy:2013, astropy:2018}, Python \citep{vanRossum95,Oliphant07}, Matplotlib \citep{Hunter07,Caswell20}, 
numpy \citep{Oliphant06,vanderWalt11} and SciPy \citep{Virtanen20,Virtanen20B}.
 
\end{acknowledgements}


\bibliographystyle{aa}
\bibliography{bpic_hill}

\newpage

\end{document}